\documentclass[fleqn,usenatbib]{mnras}

\usepackage{newtxtext,newtxmath}

\usepackage[T1]{fontenc}
\usepackage{ae,aecompl}
\usepackage{comment}
\usepackage{makecell}

\usepackage{graphicx}
\usepackage{amsmath}
\usepackage{amssymb}

\title[Fly-by encounters between two planetary systems I]{Fly-by encounters between two planetary systems I: solar system analogues}

\author[Li, Mustill \& Davies]{
Daohai Li\thanks{E-mail: li.daohai@astro.lu.se, lidaohai@gmail.com (DL)},
Alexander J. Mustill,
and Melvyn B. Davies
\\
Lund Observatory, Department of Astronomy and Theoretical Physics, Lund University, Box 43, SE-221 00 Lund, Sweden
}

\date{Accepted XXX. Received YYY; in original form ZZZ}

\pubyear{2019}

\hypersetup{draft}
\begin{document}
\label{firstpage}
\pagerange{\pageref{firstpage}--\pageref{lastpage}}
\maketitle

\begin{abstract}
Stars formed in clusters can encounter other stars at close distances. In typical 
open clusters in the Solar neighbourhood containing hundreds or thousands of member stars, 
ten to twenty per cent of Solar-mass member stars are expected to encounter another star at distances closer than 100 au. These close encounters strongly perturb the planetary systems, directly causing ejection of planets or their capture by the intruding star, as well as exciting the orbits. Using extensive $N$-body simulations, we study such fly-by encounters between two Solar System analogues, each with four giant planets from Jupiter to Neptune. We quantify the rates of loss and capture immediately after the encounter, e.g., the Neptune analogue is lost in one in four encounters within 100 au, and captured by the flying-by star in one in twelve encounters. We then perform long-term (up to 1\,Gyr) simulations investigating the ensuing post-encounter evolution. We show that large numbers of planets are removed from systems due to planet--planet interactions and that captured planets further enhance the system instability. While encounters can initially leave a planetary system containing more planets by inserting additional ones, the long-term instability causes a net reduction in planet number. A captured planet ends up on a retrograde orbit in half of the runs in which it survives for 1Gyr; also, a planet bound to its original host star but flipped during the encounter may survive. Thus, encounters between planetary systems are a channel to create counter-rotating planets, This would happen in around 1\% of systems, and such planets are potentially detectable through astrometry or direct imaging.

\end{abstract}

\begin{keywords}
celestial mechanics -- planet-star interactions -- planetary systems -- open clusters and associations: general
\end{keywords}



\section{Introduction}\label{sec-intro}

Stars often form in giant molecular clouds together with many others \citep{Lada2003} with tens to a few $10^5$ siblings \citep{PortegiesZwart2010}. Clusters with only a few tens of stars evaporate quickly due to a combination of loss of the gas, the Galactic tidal field, and internal scattering. However, those with more than $\sim$100 members can remain bound for more than 100 Myr and be seen as open clusters \citep{Adams2001a}. In the Solar neighbourhood within a few kpc, thousands of open clusters have been discovered \citep[e.g.,][]{Kharchenko2013}. Their masses, ages and sizes cover large ranges, varying by orders of magnitude. Typically, a cluster consists of hundreds of member stars within a region a few pcs \citep{Kharchenko2005,Lamers2005,Kharchenko2013}. For example, $\sim 250$ pc from the Sun, Blanco 1 is a relatively young open cluster aged $\sim100$ Myr with $\sim300$ members \citep{Kharchenko2013}. Thus the stellar number density is $\sim 100$ stars pc$^{-3}$. In this paper, we are considering encounters that will likely take place within these clusters. These clusters are formed in giant molecular clouds and are rare in number, but due to the large masses, they actually contribute a comparable total mass of stars compared to the more numerous and less massive ones \citep{Lada2003}.

In such a clustered environment, stars may encounter each other at close distances. A plethora of studies has investigated the evolution of such clusters using self-consistent $N$-body simulations. It is established that, in such clusters, typically a star experiences a few encounters within 1000 au over the course of $\sim$100 Myr \citep{Malmberg2007}. 
The combined effect of these encounters and the background ultraviolet radiation in such clusters is relatively mild and the formation of the giant planets at tens of au should not be interrupted \citep[][ and see also \citealt{Pfalzner2005,Breslau2014,Bhandare2016}]{Scally2001,Adams2006,Nicholson2019}. Actually, the occurrence rate for giant planets in the open cluster M 67 has been reported to be consistent with that of field stars \citep{Brucalassi2016, Brucalassi2017}. 


When the planets have formed, they become the target which the encounters perturb. Based on the numerical approach adopted, two broad categories of studies exist.

The first is to follow the actual dynamical evolution of the cluster in a self-consistent manner. 
Looking into clusters with hundreds to thousands of member stars,  
\citet{Scally2001,Adams2006,Malmberg2007} showed that on average, a few to 10 per cent of component stars suffer from a close encounter inside 100 au \citep{Scally2001,Proszkow2009}. However, due to mass segregation, massive 
stars, down to about a Solar-mass ($1\mathrm{M}_\odot$), are more centrally 
distributed \citep[e.g.,][]{Raboud1998,Hillenbrand1998,Parker2016}. As a result, 
stars $\gtrsim 1 \mathrm{M}_\odot$ are more prone to encounters \citep{Malmberg2011,Hands2019} 
and 10\% to 20\% of such stars witness a flyby $<$100 au. The close encounters identified within the cluster simulations could be then applied to planetary systems \citep [e.g.,][]{Vincke2016,Cai2017,Fujii2019}. Or more directly, some recent studies have been able to simultaneously evolve the planets together with the stars  \citep{Hands2019,VanElteren2019}.


The second category is a Monte-Carlo approach where encounters between the planetary system and a star or binary are created based on the properties of the assumed parent cluster. 
The evolution of a planetary system is then tracked under the effects of this encounter. 
For example, the cross-sectional areas for immediate ejection and orbital excitation for Solar System objects when they encounter binary stars have been computed \citep{Laughlin1998,Adams2001,Li2015}. 
In addition to the immediate consequences of such encounters, one can consider the long-term effects. For example, the perturbed planetary system may also become unstable hundreds of Myr after the encounter \citep{Malmberg2011,Hao2013,Davies2014}. 

The above two methods are complementary and specialise in different aspects. While the cluster method can more rigorously treat the flybys, simultaneously integrating thousands of bodies is computationally intensive. As a result, often, the length of the integration limited to $\lesssim$10 Myr \citep[e.g.,][but see also \citealt{Malmberg2007,Cai2017,Fujii2019} for longer simulation times]{Proszkow2009,Vincke2016,Hands2019}. Dedicated to single planetary systems, the Monte-Carlo approach, being less computationally demanding, allows us to create close encounters directly and to follow the evolution of the systems up to $1$ Gyr.

We build this paper on the earlier work where the authors investigated fly-by encounters between a Solar System (with four giant planets) and a single star \citep[e.g.,][]{Malmberg2011}. 
Given that exoplanets are commonly observed, with occurrence rates of a few tens of per cent \citep[e.g.,][]{Cumming2008,Cassan2012,Zhu2018}, it is natural to consider encounters where both stars host 
planetary systems.
Hence, when modelling encounters between two planetary systems, we will investigate the immediate role of the planets during the encounter and how the captured planets interact with the originals on long timescales.

The paper is organised as follows. In Section \ref{sec-enc}, we describe our encounter simulations. In Section \ref{sec-long}, we simulate the long-term evolution of the post-encounter systems up to $10^8$ or $10^9$ yr. Section \ref{sec-dis} is devoted to discussion of implications derived from this work. Finally, we summarise our main results in Section \ref{sec-con}.

\section{Encounter phase} \label{sec-enc}

\begin{figure}
\includegraphics[width=\columnwidth]{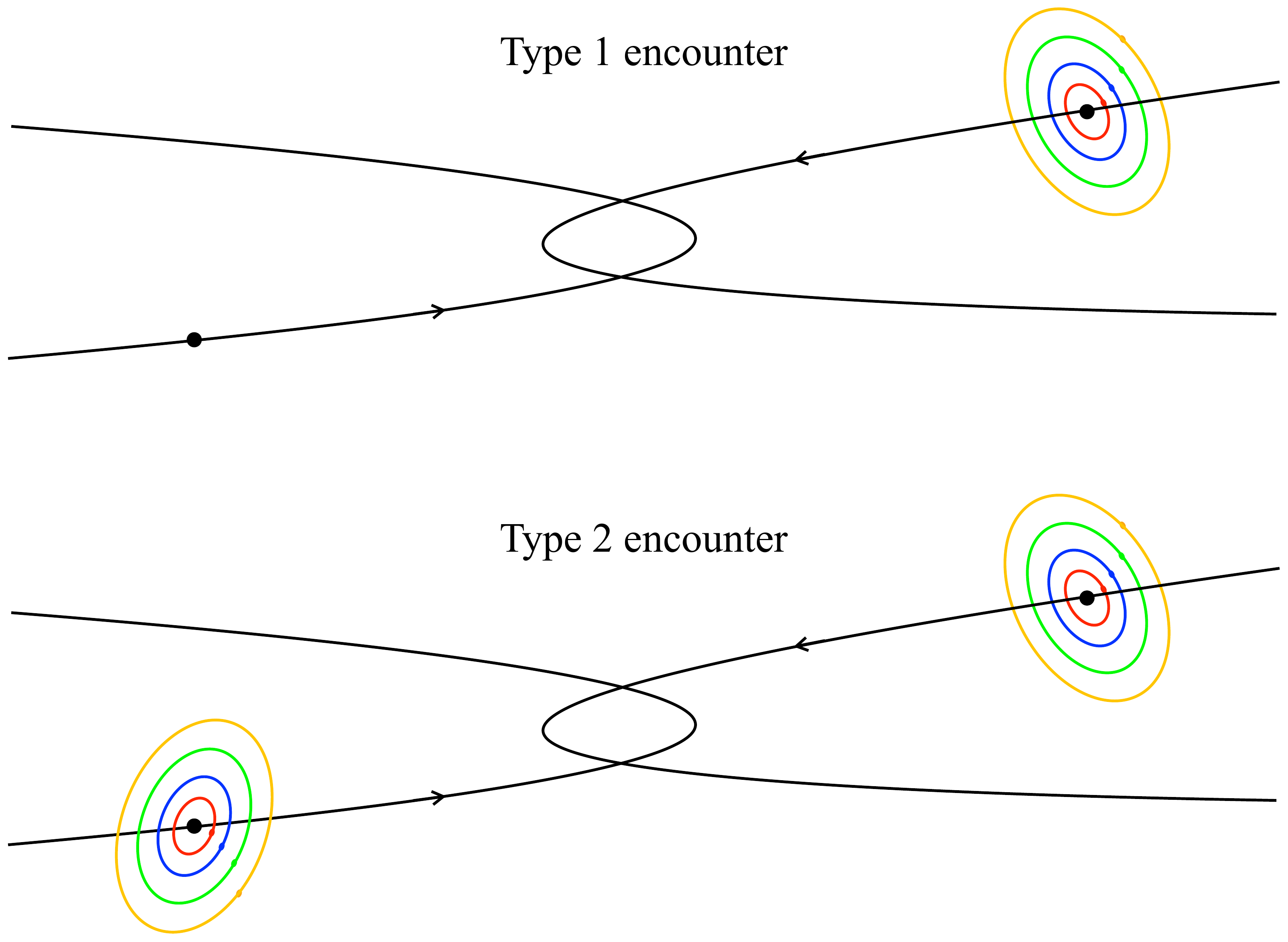}
\caption{Illustration of the two types of encounters. In Type 1, we have a single star encountering a solar system (with four giant planets) and in type 2, we study the encounter between two solar systems. Colour coding is such that red denotes Jupiter, blue for Saturn, green for Uranus and orange for Neptune.}
\label{fig-illustration-crop}
\end{figure}

We start by introducing the abbreviations. Two types of encounters are investigated in this work. In type 1 (T1), we simulate encounters between a Solar-mass star (with no planet around it) and a Solar System (with four giant planets) while in type 2 (T2), two Solar Systems are involved. See Figure \ref{fig-illustration-crop} for illustration. These are our ``encounter phase'' simulations. Because we are looking specifically into Solar-type stars in this work, all systems/stars are copies of the Solar system/the Sun. In a forthcoming paper, we will address varied planetary system configurations.

\subsection{Initial condition}

For T1 encounters, we assume the initial distance between the flying-by star and the target solar system is 2000 au, much larger than 1000 au, the usual threshold for an encounter \citep[e.g.,][]{Adams2001}. The relative velocity between the star and the barycentre of the target system is $v_\infty=$1 km s$^{-1}$, typical of young open star clusters in the solar neighbourhood \citep{Proszkow2009,Adams2010}.

The maximum encounter pericentre distance is $r_\mathrm{enc,max}=100$ au, because the emphasis of this paper is on the extreme encounters with immediate loss/capture and the long-term implications. For a given $v_\infty$ and stellar mass (one solar mass throughout this work), we consider gravitational focusing \citep[see][for instance]{Malmberg2011} and calculate a maximum impact parameter $b_\mathrm{max}$ such that $r_\mathrm{enc}=r_\mathrm{enc,max}$. Assuming that the impact parameter $b$ follows a geometrical configuration such that its probability density function (PDF)  follows $P(b)\propto b$ and the encounter is isotropic in the solid angle, we then randomly generate $b\leq b_\mathrm{max}$. At a relative velocity of $v_\infty=$1 km s$^{-1}$, gravitational focusing is significant \citep{Binney2008}. As such, a PDF for $b$ of $P(b)\propto b$ translates to one for $r_\mathrm{enc}$ as $P(r_\mathrm{enc})\propto 1$, i.e., a flat distribution in $r_\mathrm{enc}$ \citep[e.g.,][]{Malmberg2011}. Also, owing to the small $v_\infty$, such encounters are nearly parabolic with eccentricities close to unity \citep[cf.][]{Hao2013}.

For the giant planets, we acquire their ecliptic orbits from JPL Horizons \url{https://ssd.jpl.nasa.gov/?horizons} but assign random phase angles. So the target system is almost flat (Sun+4 planets) and the flying-by star is coming from a random direction.

In T2 encounters, the flying-by star is itself orbited by the four planets in their own ecliptic, assumed to be directed also randomly. Thus, the same encounter usually has two different inclinations as viewed from the ecliptics of two systems. We generate 10000 sets of initial conditions for T1 encounters and 5000 for T2 encounters. Hence, the total number of planets in the two encounter simulations is the same--40000.

Our single-value choice for $v_\infty$ may seem idealised, so we introduce another set of simulations, referred to as T1$V$. Here we follow \citet{Laughlin1998} and generate $v_\infty$ according to a Maxwellian distribution with a mean of 1 km s$^{-1}$, with other parameters the same as in the T1 simulation. We integrate 10000 runs in this simulation set.

The integration is stopped at $10^4$ yr by which time the distance between the two stars already becomes $>$1000 au and the encounter finishes. Then we store the orbital elements and state vectors for all the objects. The inclination of a planet is measured against the ecliptic plane of a star and the orbit of a flying-by star is calculated in the same reference frame.

All $N$-body simulations are carried out with {\small MERCURY} \citep{Chambers1999} using the Bulirsch-Stoer algorithm with a tolerance of  $10^{-12}$.

\subsection{Results} \label{sec-enc-res-orb}

During a close encounter, the planets in the target system may be lost from the host system and some of those can be captured by the flying-by star. We confirm previous results \citep[e.g.,][]{Malmberg2011,Pfalzner2005} that capturing a planet is only possible when $r_\mathrm{enc}$ is no larger than three times the planet's semimajor axis, showing our choice of $r_\mathrm{enc,max}=100$ au is sufficient.

Table \ref{tab-cap-ejec} lists the rates for loss and capture for each of the four planets for T1, T2 and T1$V$encounters. First, we notice that the rates for T1 and T2 encounters are almost the same. Hence, the interplanetary interaction does not have an effect on the planets' orbital stability during the encounter phase for the planetary masses considered here \citep[see also, ][]{Pfalzner2005a}. Also, the results for T1$V$ simulations with varying $v_\mathrm{\infty}$ are statistically indistinguishable from those of T1 and T2 encounters. The explanation goes as follows. For a Maxwellian distribution with a mean of 1 km s$^{-1}$ as adopted for $v_\mathrm{\infty}$ in the T1$V$ simulations, very few stars have $v_\mathrm{\infty}$ above 4 km s$^{-1}$. An encounter at $r_\mathrm{enc}$=100 au with $v_\mathrm{\infty}$= 4 km s$^{-1}$ has a velocity at closest approach of $v_\mathrm{enc}$=7 km s$^{-1}$ while one with $v_\mathrm{\infty}$= 1 km s$^{-1}$ has $v_\mathrm{enc}$=6 km s$^{-1}$. Therefore, these ``fast'' encounter have a similar amount of time to interact with the target planetary system. So not surprisingly, T1$V$ simulations show no apparent difference from those of T1 and T2 and our choice of $v_\mathrm{\infty}$=1 km s$^{-1}$ is a reasonable simplification. The low-relative-velocity encounters explored in this work have eccentricities close to unity. However, in clusters that are extremely dense and massive (orders of magnitude higher than considered in here), e.g., the Arches cluster, the encounters are highly hyperbolic with eccentricities larger than 10. As such, the planets may be disturbed to a much less extent during the encounter \citep{Olczak2012}.

\begin{table*}
\centering
\caption{Loss and capture rates for the four giant planets in type 1, 2 and 1$V$ encounters. In T1, we have a solar system encountering a single solar-type star while two solar systems are flying-by each other in T2; the relative velocity at infinity $v_\mathrm{\infty}$ for these two simulations is fixed at 1 km s$^{-1}$. In a third set of simulations, T1$V$, encounters between a solar system and a single solar-type star is investigated but with $v_\mathrm{\infty}$ randomly sampled from a Maxwellian distribution with a mean of 1 km s$^{-1}$. Note any captured planet must be lost from its original host in the first place.}
\label{tab-cap-ejec}
\begin{tabular}{ccccccc}
\hline
&\multicolumn{3}{c}{Loss (\%)} & \multicolumn{3}{c}{Capture (\%)}\\
&T1 & T2 &T1$V$ & T1 & T2&T1$V$\\
\hline
Jupiter&$4.70_{-0.39}^{+0.45}$& $4.45_{-0.37}^{+0.47}$&$4.41_{-0.35}^{+0.39}$& $1.62_{-0.20}^{+0.30}$&$1.74_{-0.25}^{+0.27}$ &$1.54_{-0.22}^{+0.32}$ \\
Saturn&$8.00_{-0.48}^{+0.57}$& $7.87_{-0.58}^{+0.44}$& $8.22_{-0.59}^{+0.54}$&$2.55_{-0.28}^{+0.31}$&$2.64_{-0.32}^{+0.30}$ &$2.90_{-0.36}^{+0.31}$ \\
Uranus&$15.59_{-0.58}^{+0.70}$& $15.55_{-0.64}^{+0.89}$& $15.83_{-0.77}^{+0.63}$&$5.28_{-0.43}^{+0.39}$&$5.36_{-0.41}^{+0.54}$ &$5.51_{-0.39}^{+0.48}$ \\
Neptune&$23.98_{-0.69}^{+0.91}$&$24.12_{-0.91}^{+0.83}$&$25.19_{-0.79}^{+0.97}$& $8.88_{-0.59}^{+0.51}$&$8.58_{-0.60}^{+0.49}$ &$8.79_{-0.42}^{+0.67}$ \\
\hline
\end{tabular}
\end{table*}

Going through each planet, we find that as expected, both highest rates occur for Neptune, the outermost planet. A quarter is lost among which a third is indeed captured by the flying-by star (9\% in total number). The inner planets are more resistant to loss and capture and these rates are linear with respect to the orbital size of a planet \citep{Li2015,Davies2014}. On the other hand, the capture-to-loss ratio of 1/3 roughly holds for all planets.

\begin{figure}
\includegraphics[width=\columnwidth]{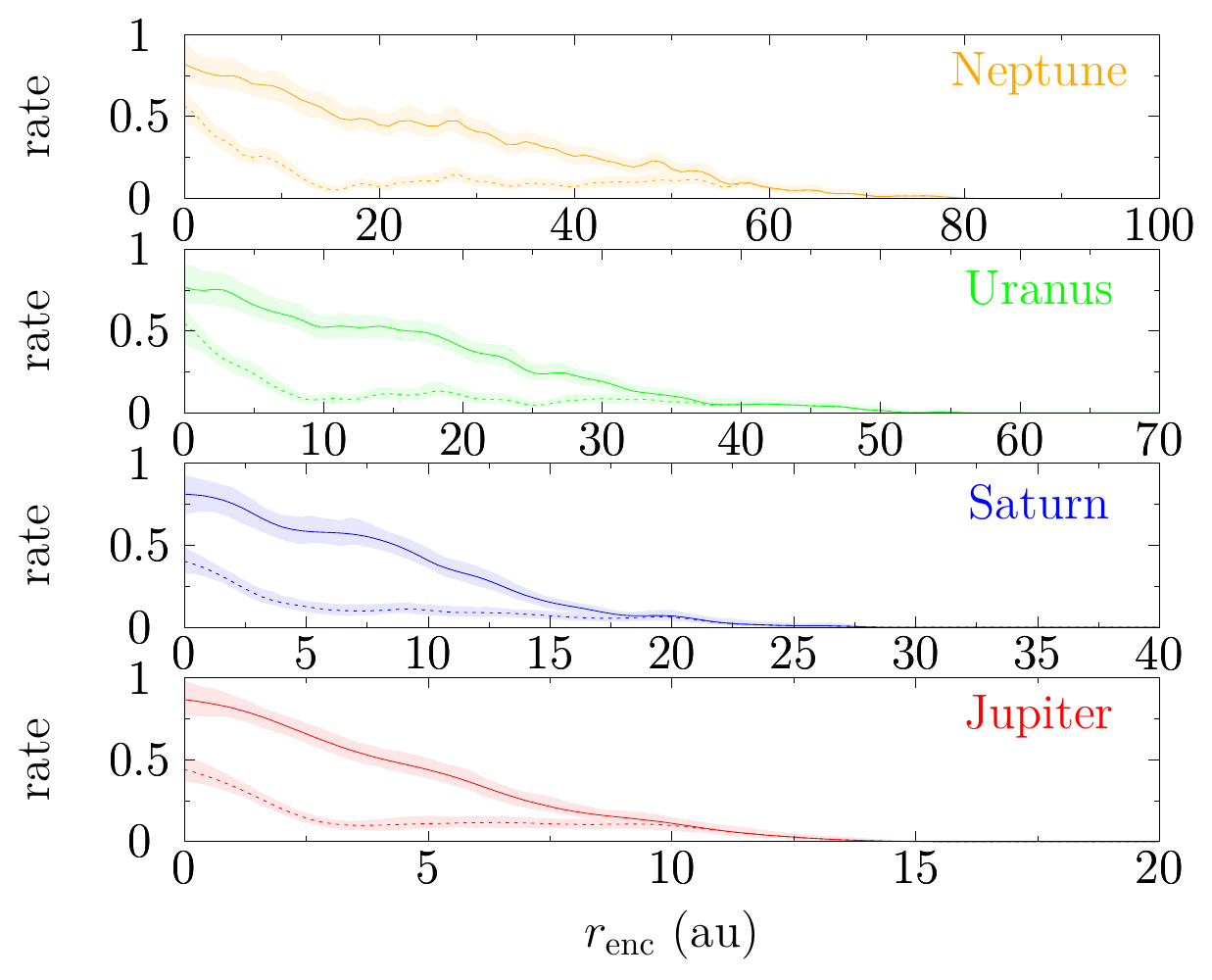}
\caption{Loss and capture rates as a function of $r_\mathrm{enc}$ for each of the planets. Solid lines are Loss rates and the dashed ones that of capture. The shaded region marks the error estimates from bootstrap resampling at 95\% confidence level. The mean capture-to-loss ratio, i.e., the quotient of the areas under the two lines, is $\sim$1/3.}
\label{fig-cap-des-effi}
\end{figure}

The chance of loss/capture depends on the geometry of the encounter \citep[e.g.,][]{Pfalzner2005,Bhandare2016,Breslau2014,Jilkova2016}. A key factor is apparently $r_\mathrm{enc}$. We plot these rates as a function of $r_\mathrm{enc}$ in Figure \ref{fig-cap-des-effi} for the four planets, all increasing at smaller encounter distances. Interestingly, the capture limit coincides with that for loss, meaning that during these distant encounters, the only way to relieve a planet of its host star is to capture it. This can be seen already, for instance, from figure 5 of \citet{Malmberg2011} where the authors showed that the maximum encounter distances for loss/capture were similar. When the encounter is deep reaching a planet $r_\mathrm{enc}\sim a_\mathrm{P}$, the loss rate is $\sim 0.5$. This rate increases to $\sim0.8$ when $r_\mathrm{enc}\ll a_\mathrm{P}$ and the capture rate rises to 0.5. When normalised against the semimajor axes, the profiles of these curves seem similar. If integrated over $r_\mathrm{enc}$, we obtain those presented in Table \ref{tab-cap-ejec}. As such, while both rates of capture and loss are complicated functions of $r_\mathrm{enc}$, their ratio averages 1/3.

\begin{figure}
\includegraphics[width=\columnwidth]{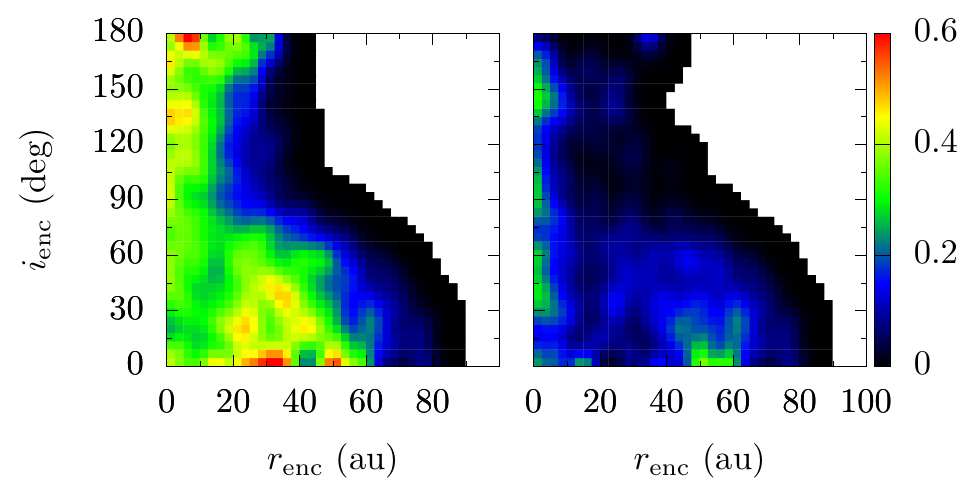}
\caption{Loss (left) and capture (right) rates as a function of encounter distance $r_\mathrm{enc}$ and $i_\mathrm{enc}$ (inclination of the encounter) for Neptune. Warmer colours mean higher chances. When integrated over $i_\mathrm{enc}$, this plot turns into Figure \ref{fig-cap-des-effi}.}
\label{fig-qi-nep}
\end{figure}

In a subtler way, the inclination of the flying-by star  $i_\mathrm{enc}$ also plays an important role in constraining the encounter geometry and thus the relative velocity between a planet and the flying-by star. In Figure \ref{fig-qi-nep} we show how loss/capture rates rely on $r_\mathrm{enc}$ and $i_\mathrm{enc}$ for Neptune. Seemingly two modes of capture, one characterised by large $r_\mathrm{enc}$ and low $i_\mathrm{enc}$ and the other by small $r_\mathrm{enc}$ with wider ranges of allowable $i_\mathrm{enc}$, emerge (and maybe some intermediate modes). This is a subset of encounters that lead to planet loss which also seem to have different modes. Prograde encounters are more effective in destabilising a planet in the target system (leading to loss or capture) at larger distances \citep[cf.][]{Bhandare2016,Jilkova2016}.

\begin{figure}
\includegraphics[width=\columnwidth]{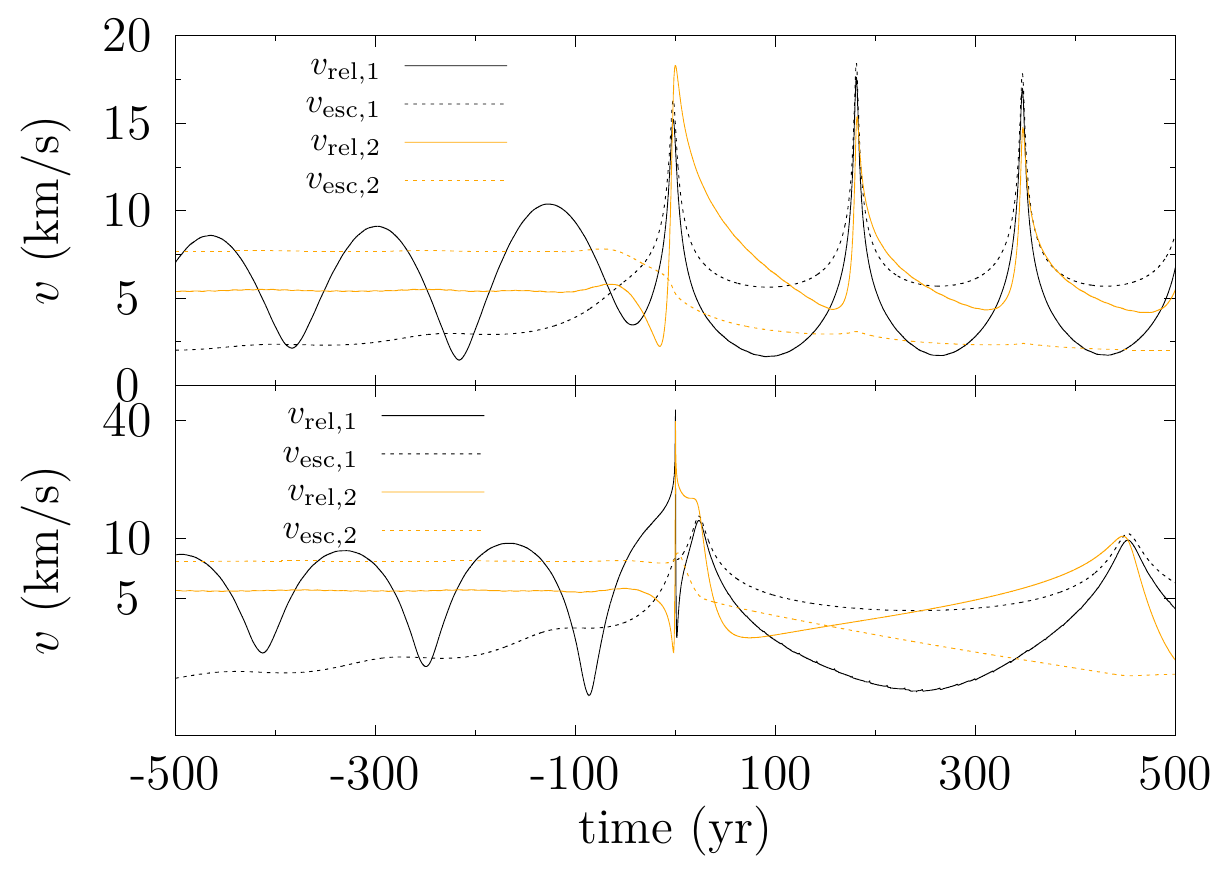}
\caption{Two capture modes (large $r_\mathrm{enc}$, top panel and small $r_\mathrm{enc}$, bottom panel) exemplified by Neptune. In both panels, the Neptune is originally orbiting star 2 (relative velocity is smaller than escape velocity for that star $v_\mathrm{esc,2}>v_\mathrm{rel,2}$) while after, both are captured by star 1 ($v_\mathrm{esc,1}>v_\mathrm{rel,1}$). See text for details.}
\label{fig-velocity-relative}
\end{figure}

We show in Figure \ref{fig-velocity-relative} the evolution of relative and escape velocities of Neptunes captured via the two modes with respect to the two stars. In both examples, Neptune is initially orbiting star 2--its relative velocity is smaller than escape velocity for that star: $v_\mathrm{rel,2}<v_\mathrm{esc,2}$. After capture, it is bound to star 1 and $v_\mathrm{rel,1}<v_\mathrm{esc,1}$.

In the top panel (large $r_\mathrm{enc}$ capture), before the closest approach at 0~yr, $v_\mathrm{esc,1}$ steadily increases, meaning that star 1 is approaching Neptune. At around -70 yr, $v_\mathrm{rel,1}<v_\mathrm{esc,1}$ but capture has not yet finished and the planet is still closer to star 2 ($v_\mathrm{esc,1}<v_\mathrm{esc,2}$). Then at -30 yr, $v_\mathrm{esc,1}$ surpasses $v_\mathrm{esc,2}$. From this time the gravitational pull of star 1 overtakes, dragging Neptune toward it. Finally at around 0 yr $v_\mathrm{esc,2}<v_\mathrm{rel,2}$.

In the bottom panel (small $r_\mathrm{enc}$ capture), things happen more drastically. Before 0 yr, $v_\mathrm{esc,2}>v_\mathrm{esc,1}$, i.e., Neptune stays closer to star 2 than to star 1. At 0 yr, the two stars are closest and star 1 begins to retreat as viewed from star 2. Now Neptune happens to be moving in the same the direction as the motion of star 1. Hence, instantly $v_\mathrm{rel,1}\ll v_\mathrm{rel,2}$ and $v_\mathrm{esc,1}> v_\mathrm{esc,2}$; soon afterwards $v_\mathrm{rel,1}< v_\mathrm{esc,1}$--capture is finished.

In addition to $r_\mathrm{enc}$ and $i_\mathrm{enc}$, other orbital parameters (e.g., the argument of pericentre) may also affect loss/capture \citep{Pfalzner2018,Pfalzner2005,Jilkova2016}.

\begin{figure}
\includegraphics[width=\columnwidth]{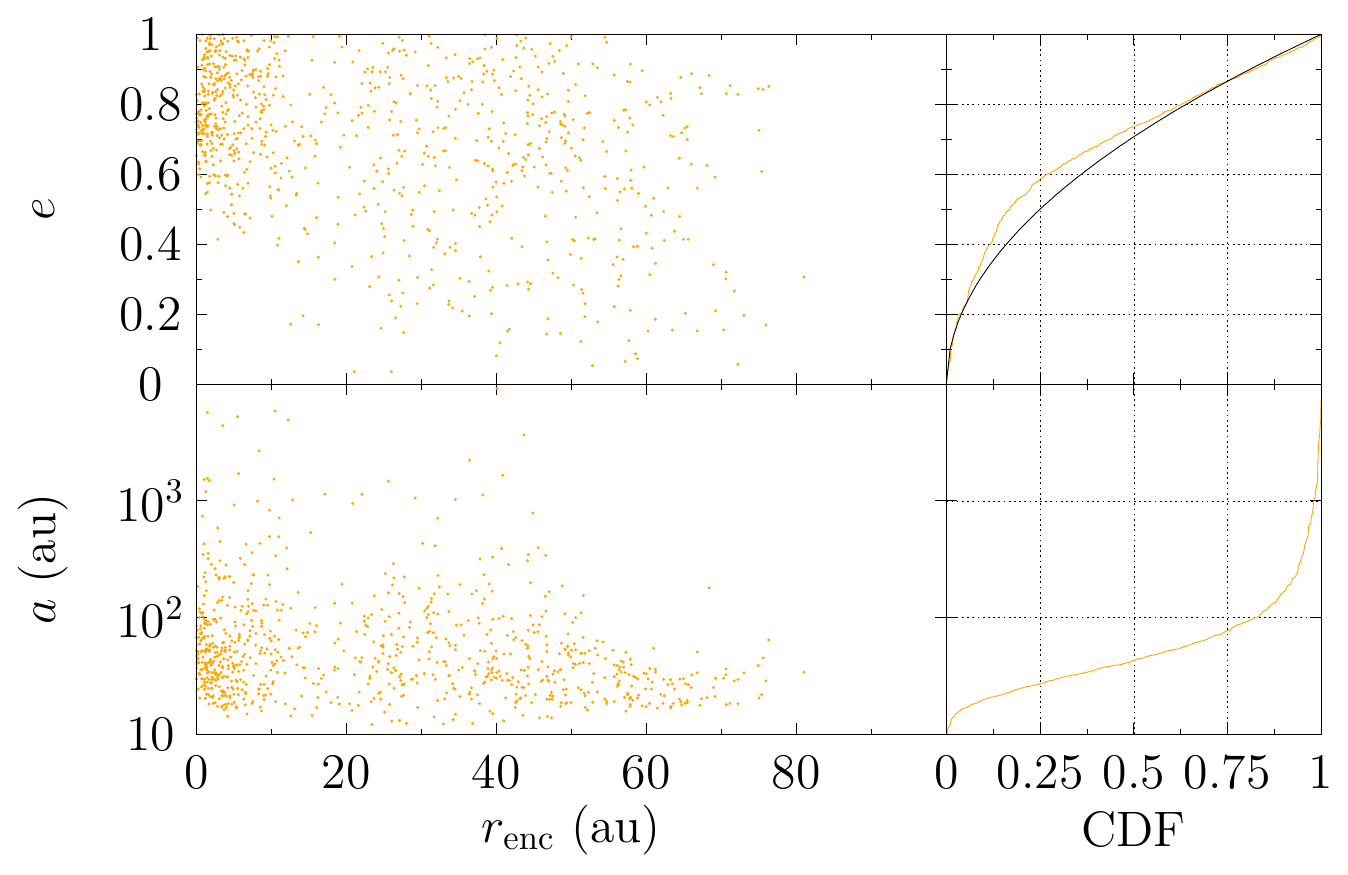}
\caption{Distribution of $e$ and $a$ for captured Neptunes immediately after encounter. On the left, we have the orbital elements $e$ and $a$ as a function of $r_\mathrm{enc}$ and on the right the CDFs are shown. The black line in the top right panel is that of thermal distribution for eccentricity, meaning CDF~$\propto e^2$.}
\label{fig-nep_enc_param}
\end{figure}

The orbits of planets captured through the two modes present different features. In Figure \ref{fig-nep_enc_param} we show the orbital elements of captured Neptunes as a function of $r_\mathrm{enc}$. At $r_\mathrm{enc}\gtrsim 50$ au, the resulting semimajor axis $a$ is usually small $<$100 au and eccentricity $e$ covers a large range from zero to unity. On the other hand, for small $r_\mathrm{enc}$ captures, $a$ may reach $\gg 100$ au, though there is a preference for small values. Meanwhile, the orbits are predominantly highly eccentric. With moderate $r_\mathrm{enc}$, $a$ and $e$ show intermediate features.

Looking at the cumulative distribution function (CDF) of the elements in the right hand side panels of Figure \ref{fig-nep_enc_param}, we find out that $\gtrsim20\%$ are with $a>100$ au. Distribution of $e$ is close to thermal but with an excess of large values, possibly due to a higher fraction of objects from small $r_\mathrm{enc}$ captures.

We do not show the distribution for inclination and they are roughly symmetric with respect to $90^\circ$ (cf. Figure \ref{fig-cap_aei_dt}). This is because we measure the orbit of a captured planet with respect to the ecliptic of the new system, assumed to be oriented randomly.

\begin{figure}
\includegraphics[width=\columnwidth]{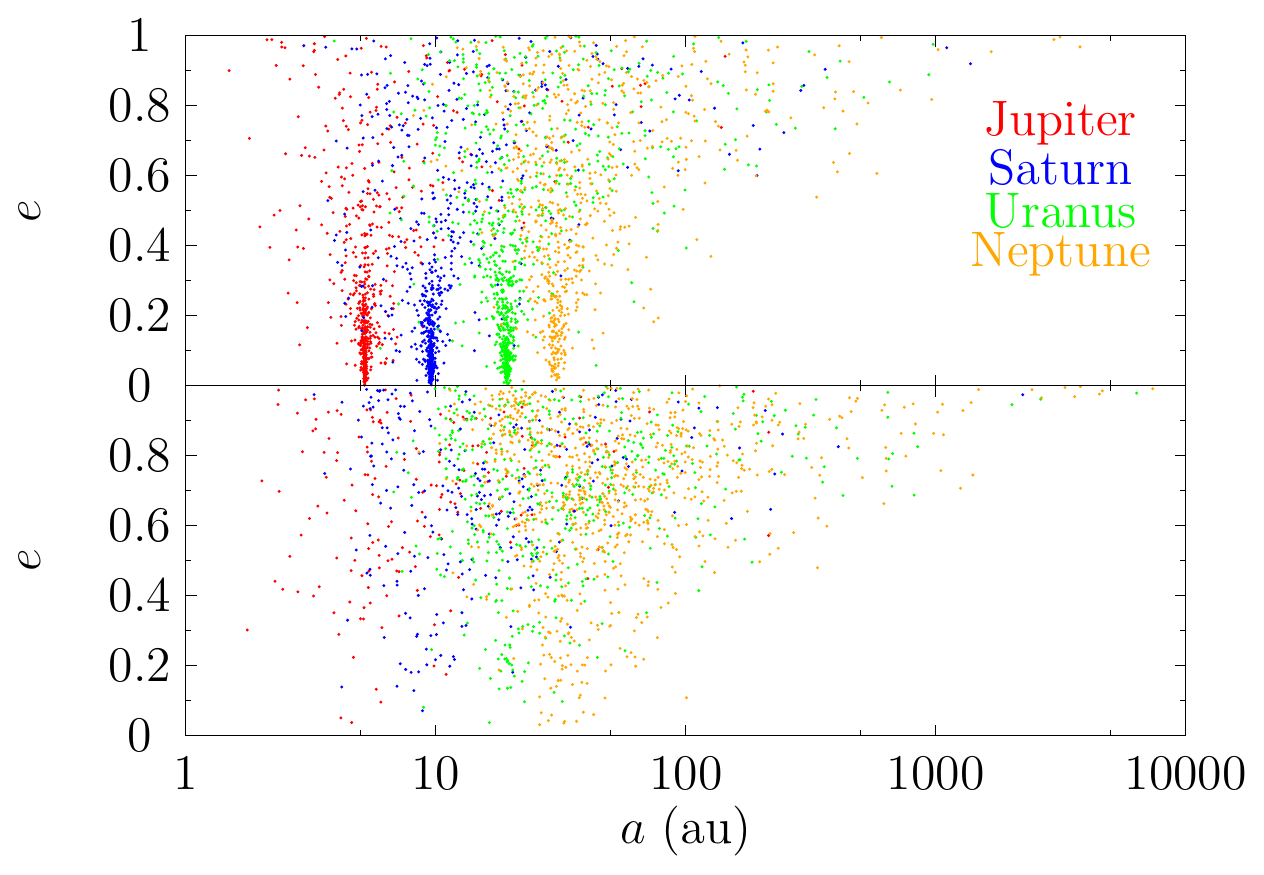}
\caption{Distribution of original (top) and captured planets (bottom) in the $(a,e)$ plane immediately after encounter.}
\label{fig-cap-ori-after}
\end{figure}

Finally, in Figure \ref{fig-cap-ori-after}, we present the orbital distribution for all the original and captured planets in the $(a,e)$ plane. The original planets are mostly moderately excited with $e$ up to 0.3 but large changes in $a$ and $e$ are also seen, overall in good agreement with existing studies \citep{Laughlin1998}. As for the captured planets, the width of the orbits captured onto is positively related to the primordial semimajor axes--small initial often leads to small captured orbits.

\section{Post encounter long-term evolution} \label{sec-long}

Having established that during the encounter phase planets play a minor role, we now proceed to investigate if their effect unfolds in the long-term post encounter evolution. We call the entire time span the post-encounter phase, as opposed to the encounter phase.

\subsection{Long-term evolution of a solar system perturbed by a flying-by star}\label{sec-general}

\subsubsection{Initial condition}

Irrespective of whether loss/capture of planets occurs, we randomly pick 1000 cases from T1 encounter simulations. The state vectors of the target star and the planets immediately after the encounter form the initial conditions for post-encounter phase simulations. Each system is integrated in isolation for $10^8$ yr. These are referred to as T1 simulations.

In a similar way, 1000 systems are chosen from T2 encounter simulations and propagated for $10^8$ yr as well. Here we omit the captured planets in T2 encounters and call these $\widetilde{\mathrm{T2}}$ simulations. Hence, these systems are only briefly perturbed by the planets in the flying-by system during the encounter phase (thus they are flying-by planets). The last section shows that the flying-by planets cannot effect significant immediate disturbance on the target system. The purpose of this $\widetilde{\mathrm{T2}}$ simulation is to examine whether these flying-by planets can have a delayed effect visible in the post-encounter long-term evolution.

We then introduce a third set of simulation where we have the same systems as in $\widetilde{\mathrm{T2}}$, the difference being that now captured planets are included. As shown in Table \ref{tab-cap-ejec}, the capture rates during these encounters are $\lesssim 10\%$. So in practice, we only need to rerun the simulations for about 100 systems, and for the remaining 900, no capture occurs and we take the results directly from $\widetilde{\mathrm{T2}}$ simulations. These are our T2 simulations.

\subsubsection{Results}
In Table \ref{tab-general-ignor} we show the survival rates at $10^4$ yr immediately after encounter and at $10^8$ yr of the T1 and $\widetilde{\mathrm{T2}}$ simulations. Being a down-sampling of those presented in Table \ref{tab-cap-ejec}, the statistics at $10^8$ yr agree in the two tables. Note here we present the survival rates whereas loss rates are shown in that table.

\begin{table}
\centering
\caption{Rates of stability of original planets immediately after encounter (IAE) and at $10^8$ yr into our long-term simulations for T2 (between two solar systems) and T1 encounters (between a solar system and a solar-mass star). In the $\widetilde{\mathrm{T2}}$ simulations, however, captured planets (if any) are omitted. Thus, the planetary systems in these simulation experience similar perturbation to those in T1 encounters. We refer to these as $\widetilde{\mathrm{T2}}$ but not T2.}
\label{tab-general-ignor}
\begin{tabular}{ccccc}
\hline
&\multicolumn{2}{c}{original} & \multicolumn{2}{c}{original}\\
&T1-IAE (\%)&T1-$10^8$ (\%)&T2-IAE (\%)&$\widetilde{\mathrm{T2}}$-$10^8$ (\%)\\
\hline
Jupiter & $95.3_{-1.5}^{+1.2}$ & $95.1_{-1.5}^{+1.3}$&$95.6_{-1.1}^{+1.3}$ & $95.6_{-1.4}^{+1.2}$\\
Saturn  & $91.4_{-1.5}^{+1.7}$ & $78.6_{-2.5}^{+2.6}$&$91.0_{-1.8}^{+1.7}$ & $76.0_{-3.2}^{+2.4}$\\
Uranus  & $84.0_{-2.3}^{+2.1}$ & $51.0_{-3.0}^{+2.9}$&$84.6_{-2.5}^{+2.4}$ & $51.6_{-3.7}^{+3.8}$\\
Neptune & $74.9_{-2.4}^{+2.7}$ & $56.6_{-2.8}^{+3.3}$&$77.2_{-2.5}^{+2.7}$ & $57.9_{-3.2}^{+3.1}$\\
\hline
\end{tabular}
\end{table}
As already been pointed out \citep{Malmberg2011,Hao2013}, instability gradually develops in the post encounter evolution and the planets are destablised severely.

Here, we observe that Uranus is the most vulnerable to instability, and is out-survived by Neptune, despite the fact that the latter is lost at higher rates during encounter phase. To be specific, while 1/3 of Uranus' loss occurs during the encounter, the remaining 2/3 gradually shows up post encounter. Notably, Saturn is characterised by a similar feature--a greater extent of destabilisation during post encounter phase than encounter phase. Both two planets have more massive inner neighbours meaning that if they gain significant eccentricity, their orbits intersect these planets and may be ejected.

For Neptune, the fractional loss during the two phases is similar. Jupiter suffers from little instability during the post-encounter phase. This is to say that the only way to effectively destabilise Jupiter is to do that during the encounter phase \citep[see also][]{Hao2013}.

The majority of the destabilised planets end up ejected (2/3), often by Jupiter \citep{Nesvorny2012}. About 1/3 dive into the Sun with heliocentric distance smaller than the sum of the planetary and solar radii whereas planetary collisions are rare due to their mutual orbital inclinations \citep[e.g.,][]{Rice2018}.

Comparing the numbers for T1 and $\widetilde{\mathrm{T2}}$ encounters, we find that the difference is negligible and we conclude that a planet that briefly flies by with its host star has little long-term influence on the target planetary system. On average, 30\% of these systems lose planets during the encounter out of which, a further 60\% lose more during the post-encounter evolution. On the other hand, out of the 70\% remaining undisrupted during the encounter, 50\% develop instability later. Hence, $\sim$60\% of the systems are damaged, either immediately during the encounter or, more likely, during the post-encounter evolution, consistent with Table \ref{tab-general-ignor}. Considering that 10-20\% of solar-type stars are expected to have experienced encounters deeper than 100 au, these results thus mean 10\% of solar system analogues formed in open clusters lose planets.

\begin{figure}
\includegraphics[width=\columnwidth]{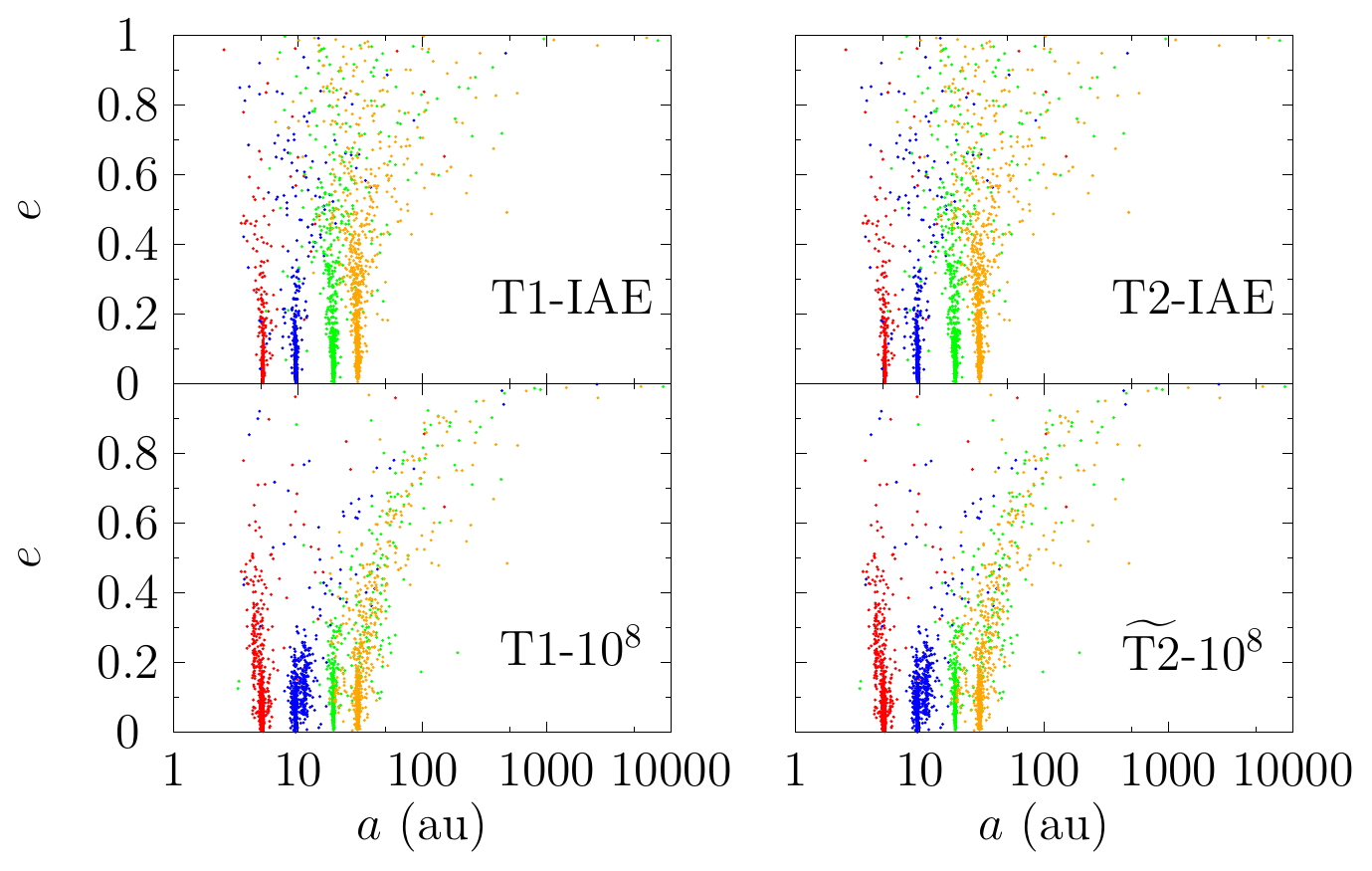}
\caption{Distribution of $a$ and $e$ for the four planets at immediately after encounter (top row, IAE) and post encounter (bottom row, at $10^8$ yr). The left column show simulation results for T1 encounters and right T2 encounters (omitting captured planets, $\widetilde{\mathrm{T2}}$). See Figure \ref{fig-illustration-crop} for colour coding.}
\label{fig-aei-post-108_two_sol_general_ignor}
\end{figure}

The distributions of $a$ and $e$ at $10^4$ yr immediately after encounter and at $10^8$ yr for T1 and $\widetilde{\mathrm{T2}}$ simulations are shown in Figure \ref{fig-aei-post-108_two_sol_general_ignor}. While those at $10^4$ yr are consistent with Figure \ref{fig-cap-ori-after}, those at $10^8$ yr roughly agrees with figure 10 of \cite{Malmberg2011} where the authors studied the long-term evolution of the solar system's giant planets after stellar encounters.

Looking at the Jupiters at at $10^8$ yr, we can identify a few subpopulations \citep[see also][]{Hao2013}, one at its starting location $a=5.2$ au with small to intermediate $e$, a group experiencing little post-encounter evolution and another at $\sim$ 4.4 au with moderate to large $e$, caused by ejecting Saturn. Another small concentration shows up at 5 au with $e$ slightly heated up, a result of the interaction with the icy planets.

Saturn develops a small pile-up at 11 au, due to the fact that it cannot eject the ice planets effectively \citep[e.g.][]{Cloutier2015} and usually transport them inward \citep{Fernandez1984,Malhotra1995}; thus it gains angular momentum and jumps outward.

In terms of eccentricity, except Jupiter, all three seemingly become colder owing to removal of high-$e$ components. Saturn, for example, is apparently eliminated severely beyond $e\gtrsim 0.3$. On the other hand, Jupiter is heated up and a large fraction achieves $e>0.3$ during the post-encounter phase, a hint of the effect of stellar encounter, as otherwise self-excitation within the planets hardly boosts Jupiter to $e>0.3$ \cite[c.f.,][]{Carrera2016}.

\begin{figure}
\includegraphics[width=\columnwidth]{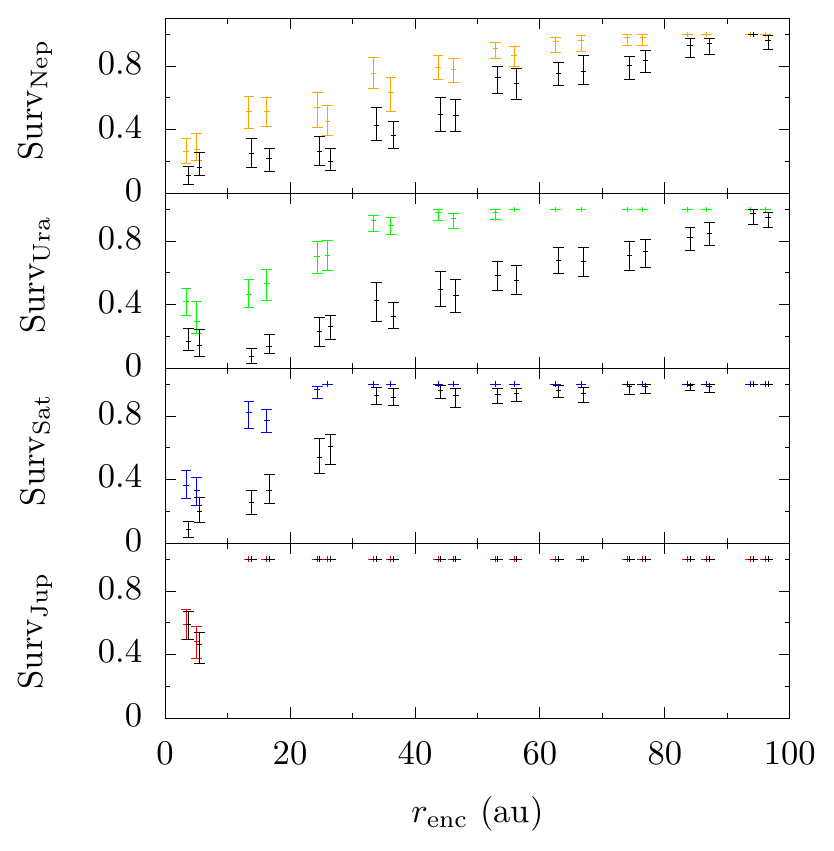}
\caption{Survival rates of planets immediately after encounter (IAE, coloured symbols) and at $10^8$ yr into our post-encounter simulations (black) as a function of $r_\mathrm{enc}$. We group the 1000 target solar systems into 10 based on encounter distance of them. Then We calculate the survival rates at IAE and at $10^8$ yr, mutually shifted by 0.4 au. These are done for T1 (right) and $\widetilde{\mathrm{T2}}$ (left) encounters, mutually displaced by 3 au. See Figure \ref{fig-illustration-crop} for colour coding.}
\label{fig-surv_e_q_two_sol}
\end{figure}

\begin{figure}
\includegraphics[width=\columnwidth]{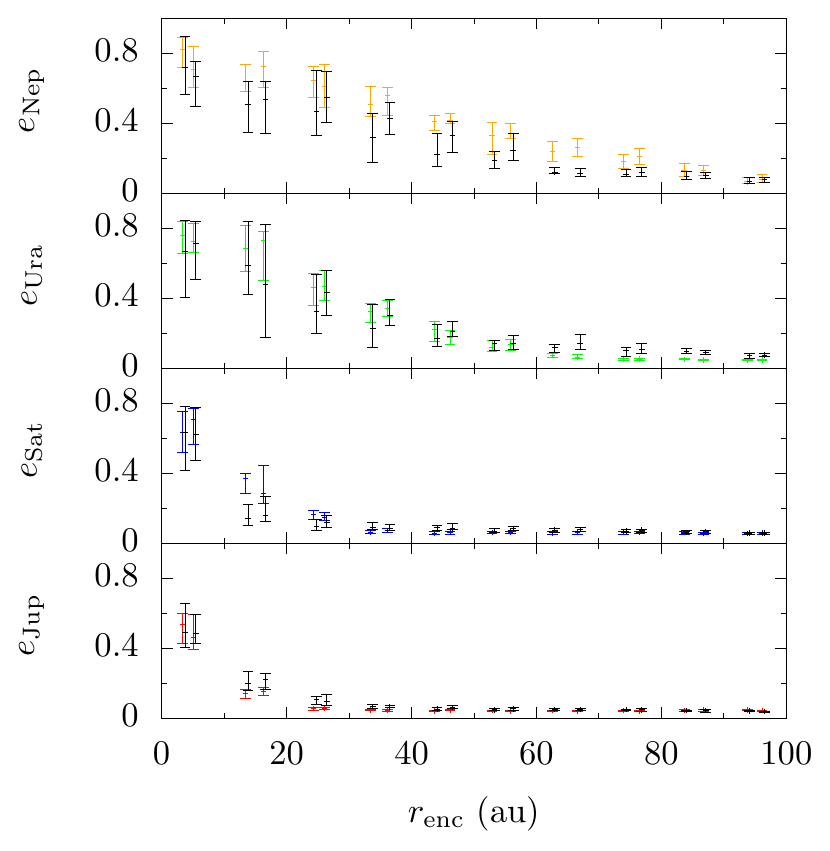}
\caption{Distribution of $e$ of the four planets immediately after encounter (coloured symbols) and at $10^8$ yr in post-encounter phase as a function of $r_\mathrm{enc}$. See Figure \ref{fig-surv_e_q_two_sol} for symbol meaning.}
\label{fig-aei_e_q_two_sol}
\end{figure}

The stability and orbital elements immediately after encounter both depend on the encounter distance during the encounter phase \citep{Spurzem2009,Pfalzner2005}. Is this information wiped out during post-encounter evolution? In Figures \ref{fig-surv_e_q_two_sol} and \ref{fig-aei_e_q_two_sol}, we show the survival rate and median eccentricity of the planets as a function of $r_\mathrm{enc}$ for the two types of encounters.

Reading from the two figures, Jupiter can only be destabilised by encounters $<$10 au and only during the encounter phase. Its eccentricity can be directly excited greatly during encounter phase by such encounters or to a smaller extent in the post-encounter phase by interplanetary interactions. In the former case, nonetheless, because the system is disrupted to a large extent, Jupiter has no planets to interact with and its $e$ is fossilised in the post-encounter phase. Jupiter cannot feel more distant encounters ($r_\mathrm{enc}>30$ au) much.

For $r_\mathrm{enc}< 10$ au, Saturn is further depleted during post-encounter phase but eccentricity distribution unchanged. A feature for Saturn is a strong depletion for moderate encounters (10 au $\lesssim r_\mathrm{enc}\lesssim 30$ au) with a drop in $e$ by $10^8$ yr because of the removal of the excited ones. Though distant encounters with $r\gtrsim30$ au cannot destabilise Saturn during encounter phase, they may induce strong interplanetary interactions that eliminate Saturn during post-encounter evolution.

The stability feature of the ice planets are similar to Saturn's: encounters at larger $r_\mathrm{enc}$ are able to cause damage post encounter though not during the encounter, Uranus being affected the most severely. The evolution of $e$ is complicated and related to the elimination of high-$e$ components. In general, Neptune becomes cooler while Uranus may be heated up.

Due to post-encounter phase interaction, destabilisation of Saturn, Uranus and Neptune is less dependent on the encounter itself.

Again, we observe no difference between T1 and $\widetilde{\mathrm{T2}}$ simulations. Can captured planets change this picture? We now proceed to examine the results of our T2 simulations, where captured planets are included during the post-encounter phase.

In Table \ref{tab-general-ori-cap}, we present the survival rates for T2 simulations. Comparing those for original planets in T1 and $\widetilde{\mathrm{T2}}$ (Table \ref{tab-general-ignor}), all original planets are more prone to instability, but only marginally and consistent with the error. Out of the 1000 systems in T2 simulations, only $\sim$100 manage to capture planets from the flying-by star. Thus, the effects of the captured planets are greatly diluted. On the other hand, captured planets themselves are removed efficiently: except Jupiter, less than half survive till the end of the simulation. Hence, in the end, we only have a few tens of captured planets left.

\begin{table}
\centering
\caption{Surviving rates of planets immediate after encounter (IAE) and post-encounter at $10^8$ yr for type 2 encounters where two solar solar systems fly by each other. Here the captured planets are included.}
\label{tab-general-ori-cap}
\begin{tabular}{ccccc}
\hline
&\multicolumn{2}{c}{original} & \multicolumn{2}{c}{captured}\\
&T2-IAE (\%)&T2-$10^8$ (\%)&T2-IAE (\%)&T2-$10^8$ (\%)\\
\hline
Jupiter&$95.6_{-1.4}^{+0.9}$ & $94.7_{-1.6}^{+1.4}$ & $1.7_{-0.6}^{+1.2}$ & $1.2_{-0.6}^{+0.7}$\\ 
Saturn &$91.0_{-1.6}^{+1.8}$ & $75.5_{-2.9}^{+2.4}$ & $2.7_{-0.9}^{+1.0}$ & $1.2_{-0.6}^{+0.7}$\\ 
Uranus &$84.6_{-2.4}^{+2.0}$ & $50.0_{-3.0}^{+3.3}$ & $5.6_{-1.5}^{+1.6}$ & $2.5_{-1.0}^{+1.1}$\\ 
Neptune&$77.2_{-2.9}^{+2.1}$ & $55.8_{-3.3}^{+2.9}$ & $9.5_{-1.7}^{+1.7}$ & $4.3_{-1.2}^{+1.3}$\\
\hline
\end{tabular}
\end{table}

\subsection{Long-term evolution of a solar system capturing one or more planets during an encounter}
In order to obtain better statistics for the long-term evolution of the captured planets, we now perform new post-encounter phase long-term simulations to $10^9$ yr.
\subsubsection{Initial condition}
From all T2 encounter phase simulations, we pick all planet-capture systems, a total of 1390. Two sets of simulations are carried out for these systems. In the first, we have both populations of captured planets (1783) and original ones (3559); this is our T2E simulations (``E'' stands for extended). In the second, only the original ones are included; this is $\widetilde{\mathrm{T2}}$E.

Here the integration time is $10^9$ yr because we want to resolve the instability at later times \citep[e.g.,][]{Hao2013}. In addition, these two sets of simulations are not to be directly compared with those in Section \ref{sec-general}, since we are now using a biased subsample of the encounters (Figure \ref{fig-qi-nep}).

\subsubsection{Results}
As before, we first count the fraction of surviving planets in Table \ref{tab-eff-cap}. Because of the bias toward close-in encounters, immediately after the encounter, we already see a far greater extent of destruction.

Looking at the original planets in T2E simulations, as analysed in Section \ref{sec-general}, a larger degree of disruption occurs during the post-encounter phase for Saturn and Uranus and this phenomenon looks more pronounced here. Similar to T2 simulations, the captured planets, except Jupiter, are also destabilised significantly and usually less than half survive, relative rates roughly consistent with Table \ref{tab-general-ori-cap}. Uranus, for example, is removed by a relative fraction of 65\% among the initial captured population. The fact that Jupiter is most resistant to interplanetary interactions agrees with previous studies \citep[e.g.,][]{Hao2013}.

It is interesting to note that the captured planets are not necessarily easier to destabilise than the originals. For example, captured Uranus out-survive its original counterpart, both by fraction and absolute number. 

Comparing T2E with $\widetilde{\mathrm{T2}}$E, we find original Jupiters are not affected much by the captured planets. The other three become significantly stabler if we turn off the captured planets. Altogether, the captured planets, though inducing a greater degree of destabilisation among the original ones, actually increase slightly the overall multiplicity both immediately after the encounter and post-encounter at $10^9$ yr. Typically, the total number of planets evolve from $3.84_{-0.06}^{+0.06}$ ($2.56_{-0.07}^{+0.07}$ originals+$1.28_{-0.03}^{+0.03}$ captives) immediately after encounter to $1.81_{-0.04}^{+0.04}$ ($1.35_{-0.04}^{+0.05}$+$0.46_{-0.03}^{+0.04}$) at $10^9$ yr. The corresponding numbers, in $\widetilde{\mathrm{T2}}$E simulations, are $2.56_{-0.07}^{+0.06}$ and $1.68_{-0.06}^{+0.05}$, respectively. So in this sense, for a given planetary system, whether it encounters a single star or another planetary system does not affect the number of planets in this system in the long term.

\begin{table}
\centering
\caption{Survival rates immediately after encounter (IAE) and $10^9$ yr into post-encounter simulations for T2 encounters. Note here we are only simulating the planet-capture systems, thus a biased sample in terms of of encounter distance whereas in Table \ref{tab-general-ori-cap}, no such bias is introduced. The last column shows simulation where no captives are considered. See text for details.}
\label{tab-eff-cap}
\begin{tabular}{cccccc}
\hline
&\multicolumn{2}{c}{captured (\%)} & \multicolumn{3}{c}{original (\%)}\\
&T2E-IAE &T2E-$10^9$ & T2E-IAE &T2E-$10^9$ &$\widetilde{\mathrm{T2}}$E-$10^9$\\
\hline
Jupiter&$12.2_{-1.6}^{+1.4}$ & $9.0_{-1.4}^{+1.6}$ & $78.8_{-2.2}^{+2.3}$ & $74.5_{-2.7}^{+2.0}$ & $78.3_{-2.2}^{+1.9}$\\
Saturn& $18.5_{-1.9}^{+2.3}$ & $7.1_{-1.1}^{+1.4}$ & $70.6_{-2.2}^{+2.9}$ & $35.1_{-2.8}^{+2.1}$ & $47.9_{-2.0}^{+3.2}$\\ 
Uranus&$37.6_{-2.5}^{+3.1}$ & $13.1_{-1.5}^{+2.5}$ & $58.0_{-3.0}^{+2.4}$ & $10.6_{-1.6}^{+1.7}$ & $28.7_{-2.7}^{+2.2}$\\ 
Neptune&$60.1_{-2.9}^{+2.4}$ & $25.3_{-2.0}^{+t2.5}$ & $48.7_{-2.5}^{+2.4}$ & $15.1_{-1.4}^{+2.1}$ & $28.4_{-2.4}^{+2.2}$\\ 
\hline
\end{tabular}
\end{table}

\begin{figure}
\includegraphics[width=\columnwidth]{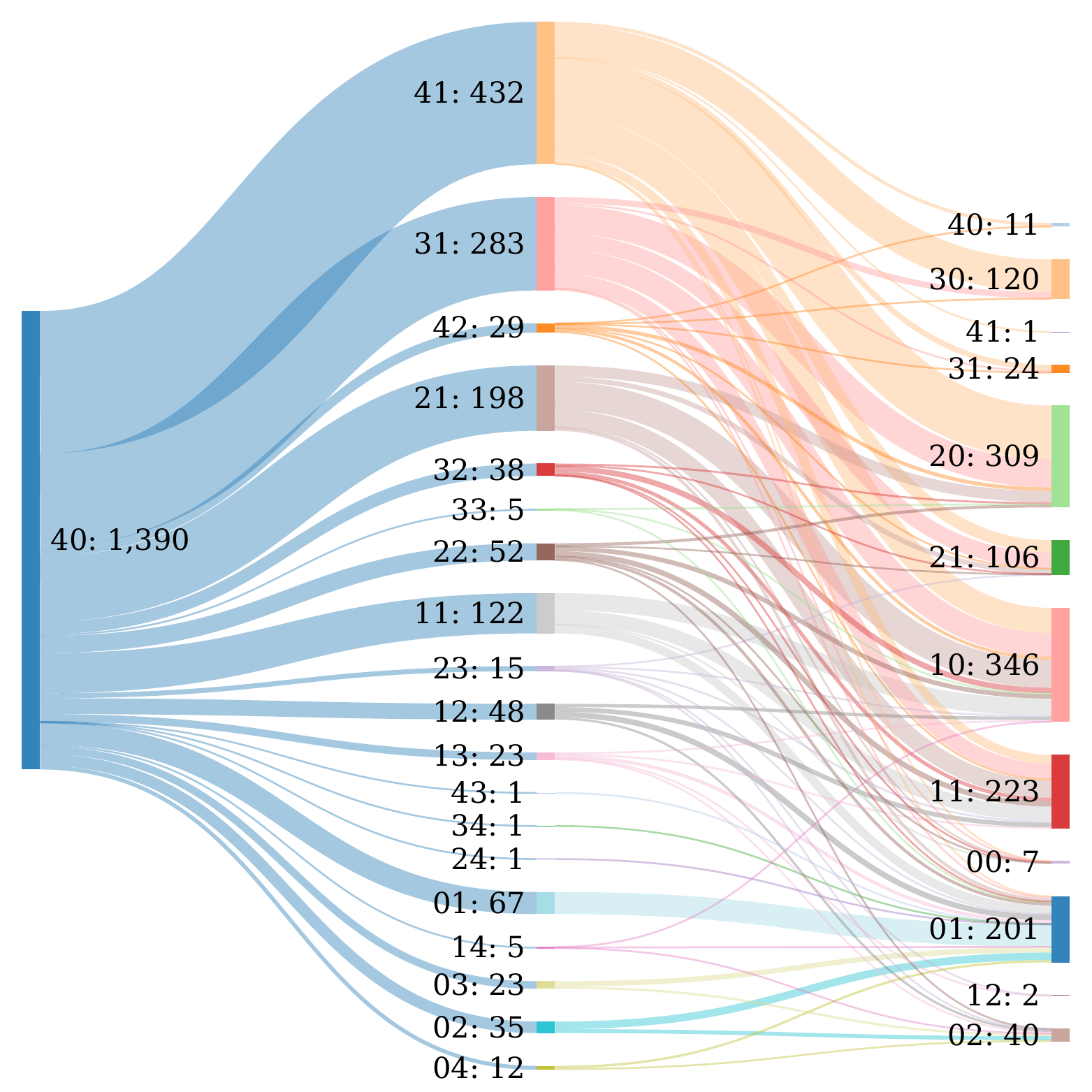}
\caption{Number of original and captured planets in the T2E simulations
  before the encounter, immediately after the encounter and $10^8$ yr into our post encounter simulations. Each label comes with two numbers. The one before the colon has two digits, the first representing the number of original planets and the second showing that of captives. The one after the colon is the number of systems observed in our simulations.}
\label{fig-numbers_shortlong}
\end{figure}

Figure \ref{fig-numbers_shortlong} illustrates how the numbers of original and captured planets evolve during and after the encounter. Here, each label comes with a two-digit number (the first digit being the number of originals and the second that of captives) followed by, after the colon, another showing the number of cases in our simulations. Before the encounter, all systems have four original planets and no captives, being ``40''. During the encounter, all acquire at least one planet from the flying-by solar system (because we are only interested in these systems here). Most frequently, for $432/1390\sim 1/3$ of the cases, one planet is captured without removing any of the four originals. The captive is predominantly Neptune and sometimes Uranus because capturing these two is possible at large encounter distances, without disrupting the original planets. Also quite often, one original planet is ejected on capturing another (``31'') and this is observed 283 times. Not surprisingly, ``21'' and ``11'' come next, with 198 and 122 instances, respectively. We note that as shown in Section \ref{sec-enc}, to retain an original planet and to capture a planet from the flying-by system are two independent processes. Hence, in all ``$n$1'' cases, $n$ being 1,2,3 or 4, the captured one is, in descending order of likelihood, Neptune, Uranus, Saturn or Jupiter. More than one planet can be captured and notably in 19 cases, all four planets hop from the one system to the other. And we point out that it is not rare to replace all original planets with captives, resulting in ``0$n$'' ($n=$ 1,2,3 or 4) systems immediately after the encounter.

As we have discussed before, the number of planets decreases dramatically during the post-encounter phase and 90\% of the 1390 systems lose at least one planet. For example, only one system of ``41'' out of the 432 cases immediately after the encounter is able to keep all the five planets throughout the $10^9$ yr post-encounter simulation. In this specific example, a Neptune is captured onto a wide orbit with $a\sim300$ au during a $r_\mathrm{enc}=\sim 50$ au encounter. The original planets are almost not affected by the encounter while the captured Neptune, due to its small mass and large orbit, is unable to disrupt the originals \citep[see, e.g.,][]{Innanen1997}. This is just one special case and a vast majority of such systems lose the captured planet during the long term evolution, evolving to ``$n$0'' and actually only 11 systems managing to keep all originals ($n=4$). This often occurs when the captive is Neptune or Uranus due to their small masses and thus vulnerability to instability. When Jupiter is captured, it likely survives to the end. 

Finally, we have look at the planet numbers at the end of our post encounter simulation. Agreeing with Table \ref{tab-eff-cap}, Figure \ref{fig-numbers_shortlong} shows that losing the captured planets is the norm and more than half of the systems end up without a captive. Not surprisingly, the most common outcome is ``10'', totalling 346. Among these, 302 are left with Jupiter only, usually on highly eccentric orbits (median eccentricity 0.35). A captured planet, if kept during the post-encounter phase, may coexist with an original, often Jupiter accompanied by a captured ice planet on a well-separated orbit, forming a hierarchical system. Also, we observe over 200 systems that end up with one or more rarely two, captured planets, accounting for 20\% of all systems.

\begin{figure}
\includegraphics[width=\columnwidth]{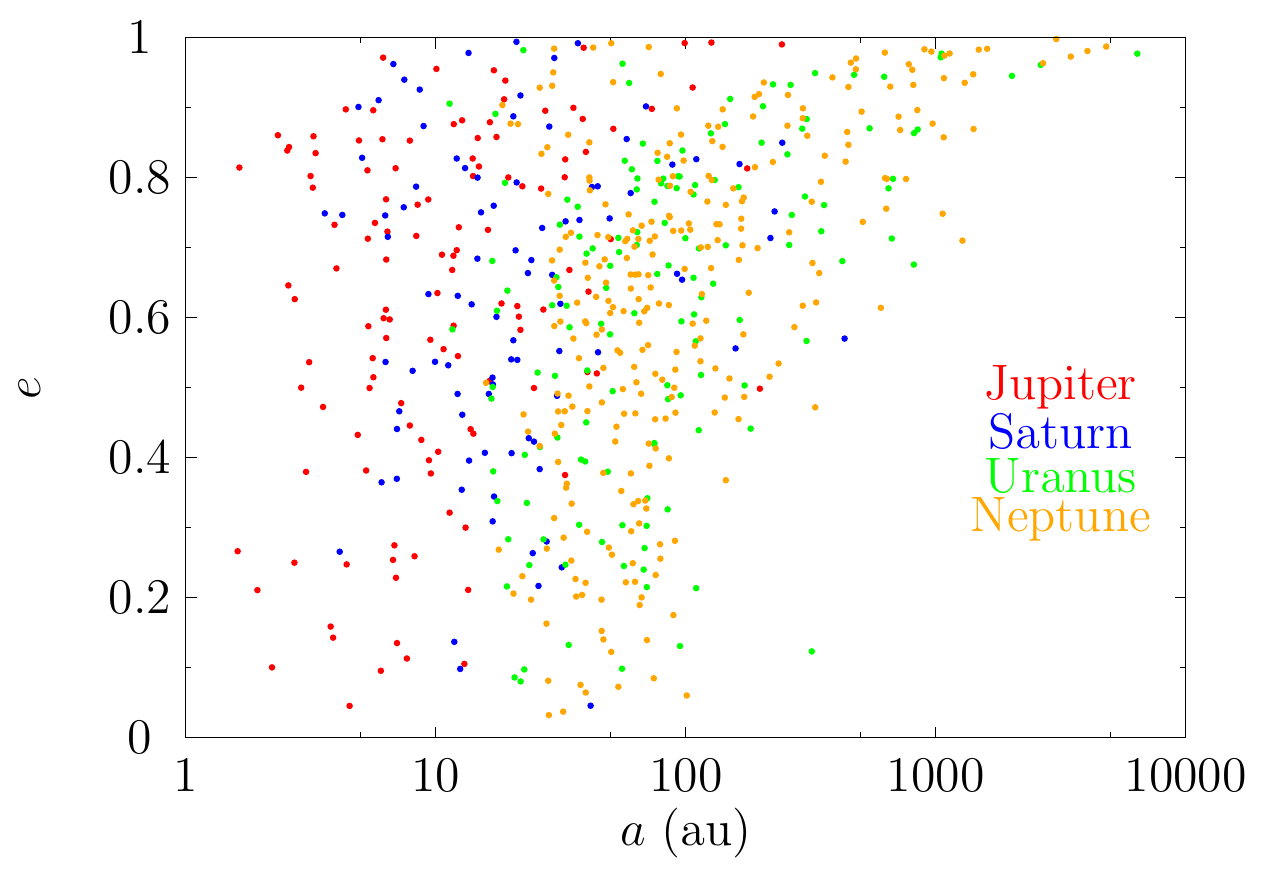}
\caption{Distribution of captured planets planets in $(a,e)$ plane at $10^9$ yr into post encounter phase.}
\label{fig-cap-ori-long}
\end{figure}

Now we turn to orbital features of the planets. Because the T2E simulations consider all captured planets, we can compare their distribution immediately after encounter (bottom panel of Figure \ref{fig-cap-ori-after}) and at $10^9$ yr into post-encounter phase (Figure \ref{fig-cap-ori-long}). We see clearly less presence of captured Neptune and Uranus in the inner solar system. Also, it appears that those captured onto wider orbits $>$100 au are resilient to later disruption.

\begin{figure}
\includegraphics[width=\columnwidth]{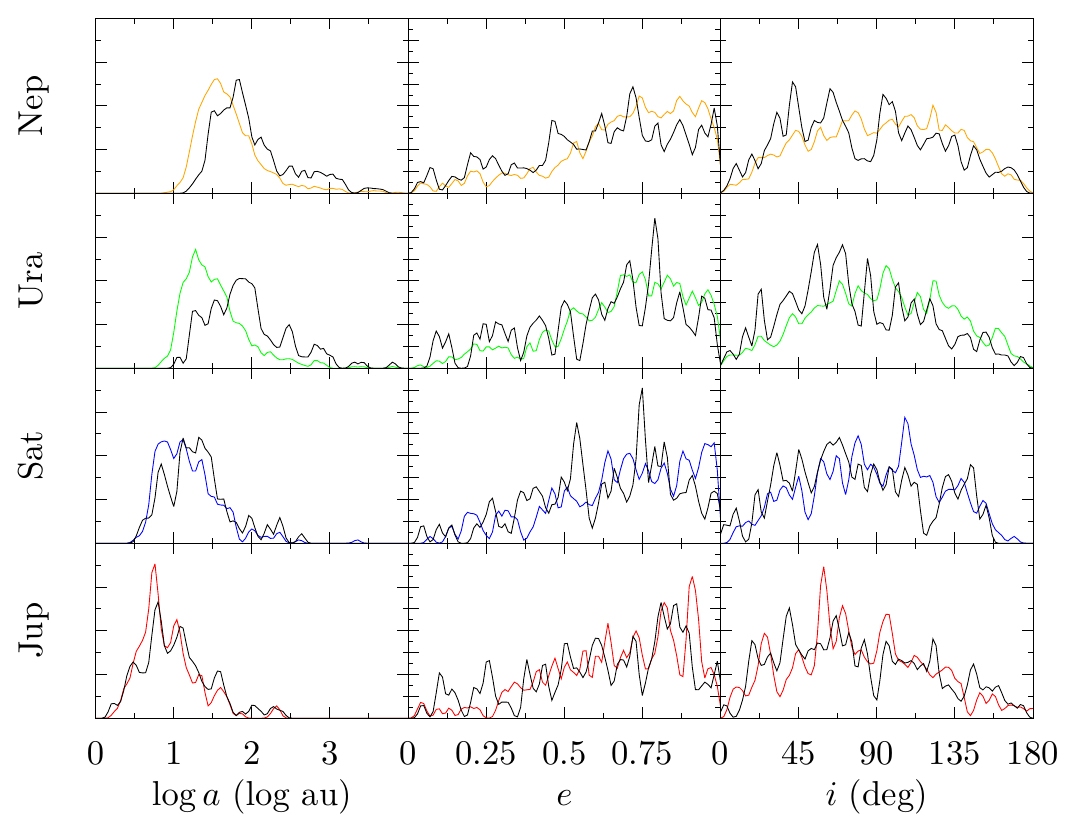}
\caption{Probability density function (PDF) of orbital elements of the captured planets. From top to bottom, the rows show Neptune, Uranus, Saturn and Jupiter' the columns, from left to right, show semimajor axis, eccentricity and inclination. Coloured lines show those immediately after the encounter and black lines at $10^8$ yr in post-encounter phase. These PDFs are normalised such that the areas under the lines are 1; thus $y$-ticks are not shown.}
\label{fig-cap_aei_dt}
\end{figure}

The probability density distribution (PDF) of the orbits of the captured planets in T2E immediately after the encounter and at $10^9$ yr are shown in Figure~\ref{fig-cap_aei_dt}.

We find that immediately after the encounter, $a$ of a captured planet is positively related to its original orbit. For instance, a captured Neptune usually obtains wider orbits than its Uranian counterpart (see also Figure \ref{fig-cap-ori-after}). The PDFs of $e$ are largely linear with respect to $e$. As expected, $i$ is isotropic upon capture. The distributions for the planets share similar shapes for each element.
 
During the post-encounter phase, the PDFs for $e$ and $i$ remain, by and large, unchanged. Those with extreme eccentricity close to unity may be preferentially eliminated and $i$ seems to shift toward prograde orbits slightly, but still nearly half survive on retrograde orbits. On the contrary, we observe systematic variations in $a$. Both ice planets show more frequent presence on wider orbits \citep{Malmberg2011}. For example, 40\% of surviving Neptunes have $a>100$ au. Indeed, these planets are totally removed inside of $10-20$ au due to strong interaction with the gas giants. That for Saturn also develops more weight in its PDF on wider orbits at $10^9$ yr owing to elimination by Jupiter. That of Jupiter has not changed much except for the emergence of a small peak at $a\sim 4.4$ au, possibly owing to ejecting Saturn; a same but more pronounced feature is seen for original Jupiter (Figure \ref{fig-cap-ori-long}).

While those on orbits hundreds of au from the Sun are free from interactions with other 
  planets in the system, they may be subject to subsequent stellar encounters. From our 
  encounter phase simulations in Section \ref{sec-enc}, a planet may be lost during an 
  encounter only if the encounter distance is comparable to its semimajor axis 
  \citep[see also, e.g.,][]{Pfalzner2005,Malmberg2011,Breslau2014,Jilkova2016}. Hence, an encounter 
  closer than $\sim$1000 au is needed to destabilise such a distant planet. An average 
  star experiences a few such encounters in a few hundred Myr \citep{Malmberg2007,Li2016a}. 
  So the planets scattered or captured onto wide orbits may be 
  vulnerable to external perturbers as long as the star cluster remains compact. 
  However, we also note that because the star cluster is dissolving over 
  time, the encounter rate drops by an order of magnitude within 100 Myr 
  \citep[e.g.,][]{Malmberg2007,Proszkow2009}. Thus, a planet captured onto a wide orbit 
  has a higher survival probability if it is captured relatively late in the 
  evolution of its host cluster. Any planets captured onto tighter orbits 
  ($<100$ au) are less vulnerable to subsequent encounters.

\section{Discussion}\label{sec-dis}
\subsection{Cross-sections at different times}
Here we have only simulated a single encounter between a Solar-mass star and a Solar System or between two Solar Systems and the long-term aftermath in a Monte-Carlo way. To put our simulation results into context, it is useful to calculate the so-called cross-sections. We define the interaction cross-section as
\begin{equation}
\sigma=\int_0^{r_\mathrm{max}} p(r)\,2\pi r \mathrm{d}r \left(1+{v^2_\mathrm{esc}(r) \over v^2_\infty} \right).
\end{equation}
Here $p(r)$ is the probability that given an encounter at distance $r$ (not impact parameter $b$), a planet is destabilised/captured (and kept) during encounter/post-encounter phases. $r_\mathrm{max}$ is the largest distance at which the two events may occur and is 100 au in our calculation. The factor inside the brackets quantify gravitational focusing \citep{Malmberg2011}, translating $r$ to impact parameter $b$; $v_\mathrm{esc}(r)$ is the escape velocity at $r$ and $v_\infty=1$ km s$^{-1}$.

We calculate four cross-sectional areas for each planet. That is destabilisation immediately after the encounter and at $10^8$ yr into post encounter phase for original planets, and capture immediately after the encounter and (stable) at $10^8$ yr for captured planets. For original planets, T1 and $\widetilde{\mathrm{T2}}$ are combined and for the captured planets, T2E is used but truncated at $10^8$ yr in order to be consistent with the originals. The results are summarised in Table \ref{tab-cross}.

We note that the values obtained here immediately after encounter are larger than those presented in \citet{Adams2006,Li2015,Laughlin1998} by factors of a few to a few tens. Their results are not directly comparable to ours because here we are calculating the cross-sectional areas exclusively for encounters between two solar-type stars, whereas the above works considered encounter involving a broad range of stellar masses, the majority being lower than the Sun. We note however, most encounters occur between unequal mass stars \citep{,Winter2018} and this affects both planet loss and capture during the encounter \citep{Bhandare2016,Jilkova2016}. In a forthcoming work, we are going to explore encounters between stars of different masses.

\begin{table}
\centering
\caption{Cross-section areas for disrupting an original planet immediately after encounter (IAE, column 2) and post encounter (at $10^8$ yr, column 3), and capture a planet at IAE (column 4) and preserving the captured planet at $10^8$ yr (column 5). For ease of comparison, the results for captured planets are taken from T2E simulations but truncated at $10^8$ yr. We note the values for Uranus and Neptune at $10^8$ yr are probably underestimated because further encounters may still induce instability during this phase as inferred from Figure \ref{fig-surv_e_q_two_sol}.}
\label{tab-cross}
\begin{tabular}{ccccc}
\hline
&\multicolumn{2}{c}{\thead{destabilisation of original \\planets ($10^4$ au$^2$)}} & \multicolumn{2}{c}{\thead{captured planets\\($10^4$ au$^2$)}}\\
&T1-IAE &T1-$10^8$ &T2-IAE & T2-$10^8$\\
\hline
Jupiter&$9.8_{-1.8}^{+2.2}$ & $10.0_{-1.8}^{+2.2}$&$4.1_{-0.8}^{+1.0}$ & $3.0_{-0.7}^{+0.8}$\\
Saturn&$19.4_{-4.0}^{+4.2}$ & $50.8_{-9.7}^{+12.8}$&$6.1_{-1.4}^{+1.9}$ & $2.4_{-0.7}^{+1.1}$\\ 
Uranus&$35.0_{-7.8}^{+10.1}$ & $110.1_{-17.9}^{+18.0}$&$12.4_{-2.6}^{+3.2}$ & $4.3_{-1.4}^{+2.2}$\\ 
Neptune&$54.3_{-12.4}^{+14.6}$ & $97.0_{-16.1}^{+17.7}$&$19.9_{-4.1}^{+5.1}$ & $8.4_{-2.5}^{+3.2}$\\ 
\hline
\end{tabular}
\end{table}

Our long-term simulations allow us to evaluate the cross-sections for the post-encounter phase. For Jupiter's destabilisation, the area is almost the same as that immediately after the encounter. In contrast, those for the other three planets increase by a factor of a few. Especially for Saturn and Uranus, the areas almost triple. These results agree with Table \ref{tab-general-ignor}.

Similarly, to keep a captured planet safe during the post-encounter phase usually has a much smaller cross-sectional area than to just capture it. For the outer three planets, the latter is a few times the former, consistent with Table \ref{tab-eff-cap}.

Exemplified by encounters between two solar-type stars, these results show that (1) for original planets, the destabilisation cross-sections can increases by factors of a few due to post-encounter phase interplanetary interactions and (2) successfully capturing a planet (hence it survives during post encounter phase) from another star is several times as infrequent as grabbing it.

We caution that our estimated cross-sectional areas at $10^8$ yr are a lower bound--as implied in Figure \ref{fig-surv_e_q_two_sol}, encounters further than 100 au may still destabilise the ice planets. And our considered simulations time of $10^8$ yr may not be enough for the instability to fully develop (for T2E, we do observe $\sim10\%$ of instability occurring after $10^8$ yr). In addition, encounters with binaries may have larger sectional areas \citep{Laughlin1998}.


\subsection{Consequences of encounters on the inner solar system}

The inner terrestrial planets are not accounted for in our simulations because the inclusion of them would require much smaller time steps. Since immediately after encounter, the destabilisation cross-sections are proportional to the orbital size of a planet \citep[see Table \ref{tab-cross} and cf.][for instance]{Li2015}, the terrestrial planets are relatively invulnerable during encounter phase \citep{Laughlin1998}. Extrapolating from Table \ref{tab-cap-ejec}, Earth only has a chance of $<$1\% of being ejected during an encounter $<$100 au. However, these less massive inner planets are exposed to stronger destabilisation later during post-encounter phase \citep{Laughlin2000}, when the giants develop instability \citep{Mustill2017,Carrera2016}. It was recently suggested that the appearance of close-in super earths might be positively correlated with that of outer cold Jupiters around solar-type stars \citep[e.g.,][]{Zhu2018a}. If so, our results would imply that these (stable) systems might not have been born in a dense cluster where encounters are frequent.

We capture a few hundreds of Jupiters in our simulations. Among these, one is captured onto a highly eccentric orbit with pericentre distance $<$0.05 au and $a=2.3$ au which is lost later due to interactions with its original equivalent. If the original Jupiter did not exist, this Jupiter would have survived and it would be circularised by tides onto a 5.3-day orbit, turning into a hot Jupiter. \citet{Brucalassi2016,Brucalassi2017} showed an excess of such planets in dense clusters. So our result points to a new formation channel, though probably less likely than some others \citep[e.g.,][]{Shara2016}. We note, as Jupiter is captured with random orbital inclinations, our model may be particularly relevant for retrograde hot Jupiters.
\subsection{Planets on wide/retrograde orbits}

Recently, a so-called Planet-Nine hundreds of au away from the Sun was proposed to explain the orbital clustering of distant Trans-Neptunian objects \citep{Batygin2016}. As previously suggested, it could be captured by the Sun from a flying-by star \citep{Mustill2016,Li2016a}. Such captured planets on wide orbits are also observed in our simulations. Additionally, we find that the original planets can be perturbed onto large-size orbits during an encounter \citep[see also][]{Malmberg2011,Li2016a}. These captured/pumped-up planets may have their pericentres beyond the inner solar system and survive the long-term planet-planet interactions. Though, in our scenario (encounter distance $r_\mathrm{enc}<100$ au), there is a good chance that the outer solar system is greatly excited \citep{Pfalzner2005}.

Tens of planetary-mass objects on $\sim100$ au orbits have been found by direct imaging and the occurrence is estimated at a few \% \citep{Ireland2011}. For example, GSC 6214-210 b is orbiting a solar mass star at 240 au and an unlikely product of planetary scattering \citep{Pearce2019}. Here our result shows that such objects can be created during stellar encounters, either scattered or captured onto \citep[see also][]{Malmberg2011} and are stable post encounter.

Many exoplanets reside on orbits significantly tilted against the equator of the host star \citep[often close-in and observed via, e.g. Rossiter-McLaughlin effect; see][]{Triaud2018}. A few tens are even rotating with projected inclinations $>90^\circ$ \citep{Breslau2019}. In our simulations, both captured and original planets can become retrograde rotators.

Our encounter and post-encounter phase simulations, combined with the rate of encounter in previous works \citep{Malmberg2007}, allow us to derive the absolute possibility for capturing retrograde planets. We do this in Equation \eqref{eq-prob} for Neptune, as the product of a set of conditional probabilities:
\begin{equation}
\label{eq-prob}
\begin{aligned} 
\centering
P_\mathrm{Neptune}=&\,\,\,\,\,\,\,P_\mathrm{encounter<100\,au} &&(15\%)
\\
&\times P_\mathrm{loss|encounter<100\,au} &&(25\%)
\\
&\times P_\mathrm{capture|loss} &&(30\%)
\\
&\times P_\mathrm{survival|capture} &&(40\%)
\\
&\times P_\mathrm{retrograde|survival} &&(50\%)
\\
&\sim0.2\%.
\end{aligned}
\end{equation}
On the other hand, the original planets can also become retrograde, mainly under the direct effect of the flying-by star during encounter \citep[see also][]{Breslau2019} but not due to the interplanetary interactions \citep[cf.][]{Chatterjee2008,Hao2013}. In our long-term post encounter simulations, we find its chance is also about 0.2\% for Neptune, the same as that for capture. This suggests that, combining both captured and original and all four planets together, the occurrence rate for retrograde planet should be $\sim$1\% around solar type stars born in open clusters.

 This rate, however, needs to be multiplied by the occurrence rate of wide-orbit gas/ice giants that is currently not well constrained. RV tells us that $\sim10\%$ of solar-type stars have giant planets within a few au \citep{Cumming2008}. Additionally, the aforementioned correlation between inner super earths with outer giant planets \citep{Zhu2018a} and the high occurrence rate of the former around solar type stars \citep{Zhu2018} seem to suggest that giant planets on wide orbits may be not uncommon.


While likely inaccessible to transit-based Rossiter-McLaughlin effect, retrograde planets created in our simulations are detectable through techniques that probe the system's motion in the plane of the sky. The reflex astrometric motion of the host star provides the full orbital geometry of a planet, up to a degeneracy in separating the ascending from the descending node of the orbital plane \citep{Perryman2014,Ranalli2018}. A ten-year Gaia mission lifetime will enable the detection of a Jupiter-mass planet at 4~au around stars out to 70~pc \citep{Ranalli2018}, with its orbital motion reliably determined up to the nodal degeneracy. Captured planets rarely end up on such tight orbits (Figure \ref{fig-cap-ori-long}),  but the original Jupiter remains in half of the systems with surviving captured planets. The captured planets may be detected by complementary direct imaging, which also determines the full orbit with the same nodal degeneracy \citep{Alzner2012}. Current direct imaging instruments such as SPHERE can detect substellar objects with a contrast ratio of $10^{-4}$ at ~0.3 arcsec \citep{Beuzit2019}, or 20~au at 70~pc, while ELT-CAM has a goal of a contrast of $5\times10^{-6}$ at 0.1 arcsec (7~au at 70~pc)\footnote{See \url{https://www.eso.org/sci/facilities/eelt/docs/ESO-193104_2_Top_Level_Requirements_for_ELT-CAM.pdf}}. Thus, even where direct imaging itself cannot discover the inner prograde planet(s), its combination with Gaia astrometry will indeed enable the identification of such retrograde systems.

\section{Conclusion}\label{sec-con}
In typical open clusters in the Solar neighbourhood with hundreds to thousands of members, a few tens of per cent of the member stars experience encounters closer than 100 au in $\sim$100 Myr \citep{Malmberg2007}. Such encounters greatly shape the planetary systems. In this work, we have performed $N$-body simulations looking into these close fly-by encounters between two solar system analogues, each carrying four giant planets. Our simulations consist of two phases: the instantaneous evolution during the encounter and the long-term evolution post-encounter. Our main findings are:
\begin{enumerate}
\item During these close encounters, a fraction, e.g., 25\% for Neptune, the outermost and most susceptible, are lost from the original host star. These lost may be either directly ejected or captured by the intruding star. For Neptune, capture occurs for 1 in 3 of the planets removed from their host, i.e., capture occurs for 8\% of Neptunes whose host experiences an encounter within 100 au.
\item Planet-planet interactions are negligible during the encounter, affecting neither loss nor capture.
\item A flying-by star, if approaching the target solar system in the same direction as the rotation of the planets, can more effectively destabilise/capture planets at further encounter distances.
\item During an encounter, a planet can be scattered/captured onto orbits orders of magnitude wider than its initial orbit. Largely decoupled from the inner system, such a planet is thus exempt from interplanetary interactions. However, these planets may be subject to subsequent stellar encounters unless they acquire wide orbits late.
\item Post-encounter, interplanetary interactions induce a great extent of planetary destruction. Less massive planets (those other than Jupiter) are especially vulnerable.
\item Except in the case of Jupiter, capturing a planet does not say much about retaining it. Using cross-sectional area estimated, we show that to keep a captured planet is several times as difficult as to capture it. 
\item A captured planet significantly increase the degree of instability. As a result, in a statistical sense, whether a flying-by star has planets (and hence can/cannot be captured) does not affect the multiplicity of the target planetary system.
\item Flybys can place wide-orbit planets onto retrograde orbits, either by capturing a planet directly onto a retrograde orbit or by flipping the orbit of an existing wide-orbit planet. In many such systems, at least one inner planet survives on a prograde orbit. Retrograde systems will soon become detectable through direct imaging supplemented by astrometry of the stellar reflex motion.
\item Combing the results obtained in this work with the occurrence rate of encounters involving solar-type stars in typical open clusters, we estimate that 10\% of Solar system analogues formed in such clusters are subject to planet loss induced by stellar encounters, either immediately during the encounter or long after it. Furthermore, 1\% of solar systems contain retrograde planets which can be either captured or original and flipped during the encounter.
\end{enumerate}

\section*{Acknowledgements}

The authors thank the anonymous referee for comments that helped improve the quality of the work. D.L. acknowledges financial support from Knut and Alice Wallenberg Foundation through two grants (2014.0017, PI: Melvyn B. Davies and 2012.0150, PI: Anders Johansen). The authors thank Alexey Bobrick at Lund Observatory for pointing us toward the creation of Figure \ref{fig-numbers_shortlong} (\url{http://sankeymatic.com/}).





\begin{thebibliography}{}
\makeatletter
\relax
\def\mn@urlcharsother{\let\do\@makeother \do\$\do\&\do\#\do\^\do\_\do\%\do\~}
\def\mn@doi{\begingroup\mn@urlcharsother \@ifnextchar [ {\mn@doi@}
  {\mn@doi@[]}}
\def\mn@doi@[#1]#2{\def\@tempa{#1}\ifx\@tempa\@empty \href
  {http://dx.doi.org/#2} {doi:#2}\else \href {http://dx.doi.org/#2} {#1}\fi
  \endgroup}
\def\mn@eprint#1#2{\mn@eprint@#1:#2::\@nil}
\def\mn@eprint@arXiv#1{\href {http://arxiv.org/abs/#1} {{\tt arXiv:#1}}}
\def\mn@eprint@dblp#1{\href {http://dblp.uni-trier.de/rec/bibtex/#1.xml}
  {dblp:#1}}
\def\mn@eprint@#1:#2:#3:#4\@nil{\def\@tempa {#1}\def\@tempb {#2}\def\@tempc
  {#3}\ifx \@tempc \@empty \let \@tempc \@tempb \let \@tempb \@tempa \fi \ifx
  \@tempb \@empty \def\@tempb {arXiv}\fi \@ifundefined
  {mn@eprint@\@tempb}{\@tempb:\@tempc}{\expandafter \expandafter \csname
  mn@eprint@\@tempb\endcsname \expandafter{\@tempc}}}

\bibitem[\protect\citeauthoryear{Adams}{Adams}{2010}]{Adams2010}
Adams F.~C.,  2010, \mn@doi [Annual Review of Astronomy and Astrophysics]
  {10.1146/annurev-astro-081309-130830}, 48, 47

\bibitem[\protect\citeauthoryear{Adams \& Laughlin}{Adams \&
  Laughlin}{2001}]{Adams2001}
Adams F.~C.,  Laughlin G.,  2001, \mn@doi [Icarus] {10.1006/icar.2000.6567},
  150, 151

\bibitem[\protect\citeauthoryear{Adams \& Myers}{Adams \&
  Myers}{2001}]{Adams2001a}
Adams F.~C.,  Myers P.~C.,  2001, \mn@doi [The Astrophysical Journal]
  {10.1086/320941}, 553, 744

\bibitem[\protect\citeauthoryear{Adams, Proszkow, Fatuzzo  \& Myers}{Adams
  et~al.}{2006}]{Adams2006}
Adams F.~C.,  Proszkow E.~M.,  Fatuzzo M.,   Myers P.~C.,  2006, \mn@doi [The
  Astrophysical Journal] {10.1086/500393}, 641, 504

\bibitem[\protect\citeauthoryear{Alzner \& Argyle}{Alzner \&
  Argyle}{2012}]{Alzner2012}
Alzner A.,  Argyle R.~W.,  2012, in Argyle R.~W.,  ed., , Observing and
  Measuring Visual Double Stars.
Springer, New York, NY, pp 71--79, \mn@doi{10.1007/978-1-4614-3945-5_7}

\bibitem[\protect\citeauthoryear{Batygin \& Brown}{Batygin \&
  Brown}{2016}]{Batygin2016}
Batygin K.,  Brown M.~E.,  2016, \mn@doi [The Astronomical Journal]
  {10.3847/0004-6256/151/2/22}, 151, 22

\bibitem[\protect\citeauthoryear{Beuzit et~al.,}{Beuzit
  et~al.}{2019}]{Beuzit2019}
Beuzit J.~L.,  et~al., 2019, (\mn@eprint {arXiv} {1902.04080})

\bibitem[\protect\citeauthoryear{Bhandare, Breslau  \& Pfalzner}{Bhandare
  et~al.}{2016}]{Bhandare2016}
Bhandare A.,  Breslau A.,   Pfalzner S.,  2016, \mn@doi [Astronomy {\&}
  Astrophysics] {10.1051/0004-6361/201628086}, 594, A53

\bibitem[\protect\citeauthoryear{Binney \& Tremaine}{Binney \&
  Tremaine}{2008}]{Binney2008}
Binney J.,  Tremaine S.,  2008, {Galactic dynamics}.
Princeton University Press, \url
  {http://adsabs.harvard.edu/abs/2008gady.book.....B}

\bibitem[\protect\citeauthoryear{Breslau \& Pfalzner}{Breslau \&
  Pfalzner}{2019}]{Breslau2019}
Breslau A.,  Pfalzner S.,  2019, \mn@doi [Astronomy {\&} Astrophysics]
  {10.1051/0004-6361/201833729}, 621, A101

\bibitem[\protect\citeauthoryear{Breslau, Steinhausen, Vincke  \&
  Pfalzner}{Breslau et~al.}{2014}]{Breslau2014}
Breslau A.,  Steinhausen M.,  Vincke K.,   Pfalzner S.,  2014, \mn@doi
  [Astronomy {\&} Astrophysics] {10.1051/0004-6361/201323043}, 565, A130

\bibitem[\protect\citeauthoryear{Brucalassi et~al.,}{Brucalassi
  et~al.}{2016}]{Brucalassi2016}
Brucalassi A.,  et~al., 2016, \mn@doi [Astronomy {\&} Astrophysics]
  {10.1051/0004-6361/201527561}, 592, L1

\bibitem[\protect\citeauthoryear{Brucalassi et~al.,}{Brucalassi
  et~al.}{2017}]{Brucalassi2017}
Brucalassi A.,  et~al., 2017, \mn@doi [Astronomy {\&} Astrophysics]
  {10.1051/0004-6361/201527562}, 603, A85

\bibitem[\protect\citeauthoryear{Cai, Kouwenhoven, Zwart  \& Spurzem}{Cai
  et~al.}{2017}]{Cai2017}
Cai M.~X.,  Kouwenhoven M. B.~N.,  Zwart S. F.~P.,   Spurzem R.,  2017, \mn@doi
  [Monthly Notices of the Royal Astronomical Society] {10.1093/mnras/stx1464},
  470, 4337

\bibitem[\protect\citeauthoryear{Carrera, Davies  \& Johansen}{Carrera
  et~al.}{2016}]{Carrera2016}
Carrera D.,  Davies M.~B.,   Johansen A.,  2016, \mn@doi [Monthly Notices of
  the Royal Astronomical Society] {10.1093/mnras/stw2218}, 463, 3226

\bibitem[\protect\citeauthoryear{Cassan et~al.,}{Cassan
  et~al.}{2012}]{Cassan2012}
Cassan A.,  et~al., 2012, \mn@doi [Nature] {10.1038/nature10684}, 481, 167

\bibitem[\protect\citeauthoryear{Chambers}{Chambers}{1999}]{Chambers1999}
Chambers J.~E.,  1999, \mn@doi [Monthly Notices of the Royal Astronomical
  Society] {10.1046/j.1365-8711.1999.02379.x}, 304, 793

\bibitem[\protect\citeauthoryear{Chatterjee, Ford, Matsumura  \&
  Rasio}{Chatterjee et~al.}{2008}]{Chatterjee2008}
Chatterjee S.,  Ford E.~B.,  Matsumura S.,   Rasio F.~a.,  2008, \mn@doi [The
  Astrophysical Journal] {10.1086/590227}, 686, 580

\bibitem[\protect\citeauthoryear{Cloutier, Tamayo  \& Valencia}{Cloutier
  et~al.}{2015}]{Cloutier2015}
Cloutier R.,  Tamayo D.,   Valencia D.,  2015, \mn@doi [The Astrophysical
  Journal] {10.1088/0004-637X/813/1/8}, 813, 8

\bibitem[\protect\citeauthoryear{Cumming, Butler, Marcy, Vogt, Wright  \&
  Fischer}{Cumming et~al.}{2008}]{Cumming2008}
Cumming A.,  Butler R.~P.,  Marcy G.~W.,  Vogt S.~S.,  Wright J.~T.,   Fischer
  D.~A.,  2008, \mn@doi [Publications of the Astronomical Society of the
  Pacific] {10.1086/588487}, 120, 531

\bibitem[\protect\citeauthoryear{Davies, Adams, Armitage, Chambers, Ford,
  Morbidelli, Raymond  \& Veras}{Davies et~al.}{2014}]{Davies2014}
Davies M.~B.,  Adams F.~C.,  Armitage P.,  Chambers J.,  Ford E.,  Morbidelli
  A.,  Raymond S.~N.,   Veras D.,  2014, in , Protostars and Planets VI.
p.~23, \mn@doi{10.2458/azu_uapress_9780816531240-ch034}

\bibitem[\protect\citeauthoryear{Fern{\'{a}}ndez \& Ip}{Fern{\'{a}}ndez \&
  Ip}{1984}]{Fernandez1984}
Fern{\'{a}}ndez J.~A.,  Ip W.~H.,  1984, \mn@doi [Icarus]
  {10.1016/0019-1035(84)90101-5}, 58, 109

\bibitem[\protect\citeauthoryear{Fujii \& Hori}{Fujii \&
  Hori}{2019}]{Fujii2019}
Fujii M.,  Hori Y.,  2019, \mn@doi [Astronomy {\&} Astrophysics]
  {10.1051/0004-6361/201834677}

\bibitem[\protect\citeauthoryear{Hands, Dehnen, Gration, Stadel  \&
  Moore}{Hands et~al.}{2019}]{Hands2019}
Hands T.~O.,  Dehnen W.,  Gration A.,  Stadel J.,   Moore B.,  2019, MNRAS,
  000, 1

\bibitem[\protect\citeauthoryear{Hao, Kouwenhoven  \& Spurzem}{Hao
  et~al.}{2013}]{Hao2013}
Hao W.,  Kouwenhoven M.~B.,   Spurzem R.,  2013, \mn@doi [Monthly Notices of
  the Royal Astronomical Society] {10.1093/mnras/stt771}, 433, 867

\bibitem[\protect\citeauthoryear{Hillenbrand \& Hartmann}{Hillenbrand \&
  Hartmann}{1998}]{Hillenbrand1998}
Hillenbrand L.~A.,  Hartmann L.~W.,  1998, \mn@doi [The Astrophysical Journal]
  {10.1086/305076}, 492, 540

\bibitem[\protect\citeauthoryear{Innanen, Zheng, Mikkola  \& Valtonen}{Innanen
  et~al.}{1997}]{Innanen1997}
Innanen K.~a.,  Zheng J.~Q.,  Mikkola S.,   Valtonen M.~J.,  1997, \mn@doi [The
  Astronomical Journal] {10.1086/118405}, 113, 1915

\bibitem[\protect\citeauthoryear{Ireland, Kraus, Martinache, Law  \&
  Hillenbrand}{Ireland et~al.}{2011}]{Ireland2011}
Ireland M.~J.,  Kraus A.,  Martinache F.,  Law N.,   Hillenbrand L.~A.,  2011,
  \mn@doi [Astrophysical Journal] {10.1088/0004-637X/726/2/113}, 726, 113

\bibitem[\protect\citeauthoryear{J{\'{i}}lkov{\'{a}}, Hamers, Hammer  \&
  Zwart}{J{\'{i}}lkov{\'{a}} et~al.}{2016}]{Jilkova2016}
J{\'{i}}lkov{\'{a}} L.,  Hamers A.~S.,  Hammer M.,   Zwart S.~P.,  2016,
  \mn@doi [Monthly Notices of the Royal Astronomical Society]
  {10.1093/mnras/stw264}, 457, 4218

\bibitem[\protect\citeauthoryear{Kharchenko, Piskunov, R{\"{o}}ser, Schilbach
  \& Scholz}{Kharchenko et~al.}{2005}]{Kharchenko2005}
Kharchenko N.~V.,  Piskunov A.~E.,  R{\"{o}}ser S.,  Schilbach E.,   Scholz
  R.-D.,  2005, \mn@doi [Astronomy {\&} Astrophysics]
  {10.1051/0004-6361:20042523}, 438, 1163

\bibitem[\protect\citeauthoryear{Kharchenko, Piskunov, Schilbach, R{\"{o}}ser
  \& Scholz}{Kharchenko et~al.}{2013}]{Kharchenko2013}
Kharchenko N.~V.,  Piskunov A.~E.,  Schilbach E.,  R{\"{o}}ser S.,   Scholz
  R.-D.,  2013, \mn@doi [Astronomy {\&} Astrophysics]
  {10.1051/0004-6361/201322302}, 558, A53

\bibitem[\protect\citeauthoryear{Lada \& Lada}{Lada \& Lada}{2003}]{Lada2003}
Lada C.~J.,  Lada E.~A.,  2003, \mn@doi [Annual Review of Astronomy and
  Astrophysics] {10.1146/annurev.astro.41.011802.094844}, 41, 57

\bibitem[\protect\citeauthoryear{Lamers, Gieles, Bastian, Baumgardt, Kharchenko
   \& {Portegies Zwart}}{Lamers et~al.}{2005}]{Lamers2005}
Lamers H. J. G. L.~M.,  Gieles M.,  Bastian N.,  Baumgardt H.,  Kharchenko
  N.~V.,   {Portegies Zwart} S.,  2005, \mn@doi [Astronomy {\&} Astrophysics]
  {10.1051/0004-6361:20042241}, 441, 117

\bibitem[\protect\citeauthoryear{Laughlin \& Adams}{Laughlin \&
  Adams}{1998}]{Laughlin1998}
Laughlin G.,  Adams F.,  1998, \mn@doi [The Astrophysical Journal]
  {10.1086/311736}, 508, L171

\bibitem[\protect\citeauthoryear{Laughlin \& Adams}{Laughlin \&
  Adams}{2000}]{Laughlin2000}
Laughlin G.,  Adams F.~C.,  2000, \mn@doi [Icarus] {10.1006/icar.2000.6355},
  145, 614

\bibitem[\protect\citeauthoryear{Li \& Adams}{Li \& Adams}{2015}]{Li2015}
Li G.,  Adams F.~C.,  2015, \mn@doi [Monthly Notices of the Royal Astronomical
  Society] {10.1093/mnras/stv012}, 448, 344

\bibitem[\protect\citeauthoryear{Li \& Adams}{Li \& Adams}{2016}]{Li2016a}
Li G.,  Adams F.~C.,  2016, \mn@doi [The Astrophysical Journal]
  {10.3847/2041-8205/823/1/L3}, 823, L3

\bibitem[\protect\citeauthoryear{Malhotra}{Malhotra}{1995}]{Malhotra1995}
Malhotra R.,  1995, \mn@doi [The Astronomical Journal] {10.1086/117532}, 110,
  14

\bibitem[\protect\citeauthoryear{Malmberg, {De Angeli}, Davies, Church, MacKey
  \& Wilkinson}{Malmberg et~al.}{2007}]{Malmberg2007}
Malmberg D.,  {De Angeli} F.,  Davies M.~B.,  Church R.~P.,  MacKey D.,
  Wilkinson M.~I.,  2007, \mn@doi [Monthly Notices of the Royal Astronomical
  Society] {10.1111/j.1365-2966.2007.11885.x}, 378, 1207

\bibitem[\protect\citeauthoryear{Malmberg, Davies  \& Heggie}{Malmberg
  et~al.}{2011}]{Malmberg2011}
Malmberg D.,  Davies M.~B.,   Heggie D.~C.,  2011, \mn@doi [Monthly Notices of
  the Royal Astronomical Society] {10.1111/j.1365-2966.2010.17730.x}, 411, 859

\bibitem[\protect\citeauthoryear{Mustill, Raymond  \& Davies}{Mustill
  et~al.}{2016}]{Mustill2016}
Mustill A.~J.,  Raymond S.~N.,   Davies M.~B.,  2016, \mn@doi [Monthly Notices
  of the Royal Astronomical Society: Letters] {10.1093/mnrasl/slw075}, 460,
  L109

\bibitem[\protect\citeauthoryear{Mustill, Davies  \& Johansen}{Mustill
  et~al.}{2017}]{Mustill2017}
Mustill A.~J.,  Davies M.~B.,   Johansen A.,  2017, \mn@doi [Monthly Notices of
  the Royal Astronomical Society] {10.1093/mnras/stx693}, 468, 3000

\bibitem[\protect\citeauthoryear{Nesvorn{\'{y}} \& Morbidelli}{Nesvorn{\'{y}}
  \& Morbidelli}{2012}]{Nesvorny2012}
Nesvorn{\'{y}} D.,  Morbidelli A.,  2012, \mn@doi [The Astronomical Journal]
  {10.1088/0004-6256/144/4/117}, 144, 117

\bibitem[\protect\citeauthoryear{Nicholson, Parker, Church, Davies, Fearon  \&
  Walton}{Nicholson et~al.}{2019}]{Nicholson2019}
Nicholson R.~B.,  Parker R.~J.,  Church R.~P.,  Davies M.~B.,  Fearon N.~M.,
  Walton S. R.~J.,  2019, \mn@doi [Monthly Notices of the Royal Astronomical
  Society] {10.1093/mnras/stz606}, 485, 4893

\bibitem[\protect\citeauthoryear{Olczak, Kaczmarek, Harfst, Pfalzner  \&
  {Portegies Zwart}}{Olczak et~al.}{2012}]{Olczak2012}
Olczak C.,  Kaczmarek T.,  Harfst S.,  Pfalzner S.,   {Portegies Zwart} S.,
  2012, \mn@doi [The Astrophysical Journal] {10.1088/0004-637X/756/2/123}, 756,
  123

\bibitem[\protect\citeauthoryear{Parker, Goodwin, Wright, Meyer  \&
  Quanz}{Parker et~al.}{2016}]{Parker2016}
Parker R.~J.,  Goodwin S.~P.,  Wright N.~J.,  Meyer M.~R.,   Quanz S.~P.,
  2016, \mn@doi [Monthly Notices of the Royal Astronomical Society: Letters]
  {10.1093/mnrasl/slw061}, 459, L119

\bibitem[\protect\citeauthoryear{Pearce, Kraus, Dupuy, Ireland, Rizzuto,
  Bowler, Birchall  \& Wallace}{Pearce et~al.}{2019}]{Pearce2019}
Pearce L.~A.,  Kraus A.~L.,  Dupuy T.~J.,  Ireland M.~J.,  Rizzuto A.~C.,
  Bowler B.~P.,  Birchall E.~K.,   Wallace A.~L.,  2019, \mn@doi [The
  Astronomical Journal] {10.3847/1538-3881/aafacb}, 157, 71

\bibitem[\protect\citeauthoryear{Perryman, Hartman, Bakos  \&
  Lindegren}{Perryman et~al.}{2014}]{Perryman2014}
Perryman M.,  Hartman J.,  Bakos G.~{\'{A}}.,   Lindegren L.,  2014, \mn@doi
  [The Astrophysical Journal] {10.1088/0004-637X/797/1/14}, 797, 14

\bibitem[\protect\citeauthoryear{Pfalzner, Vogel, Scharw{\"{a}}chter  \&
  Olczak}{Pfalzner et~al.}{2005a}]{Pfalzner2005}
Pfalzner S.,  Vogel P.,  Scharw{\"{a}}chter J.,   Olczak C.,  2005a, \mn@doi
  [Astronomy {\&} Astrophysics] {10.1051/0004-6361:20042467}, 437, 967

\bibitem[\protect\citeauthoryear{Pfalzner, Umbreit  \& Henning}{Pfalzner
  et~al.}{2005b}]{Pfalzner2005a}
Pfalzner S.,  Umbreit S.,   Henning T.,  2005b, \mn@doi [The Astrophysical
  Journal] {10.1086/431350}, 629, 526

\bibitem[\protect\citeauthoryear{Pfalzner, Bhandare, Vincke  \&
  Lacerda}{Pfalzner et~al.}{2018}]{Pfalzner2018}
Pfalzner S.,  Bhandare A.,  Vincke K.,   Lacerda P.,  2018, \mn@doi [The
  Astrophysical Journal] {10.3847/1538-4357/aad23c}, 863, 45

\bibitem[\protect\citeauthoryear{{Portegies Zwart}, McMillan  \&
  Gieles}{{Portegies Zwart} et~al.}{2010}]{PortegiesZwart2010}
{Portegies Zwart} S.~F.,  McMillan S.~L.,   Gieles M.,  2010, \mn@doi [Annual
  Review of Astronomy and Astrophysics] {10.1146/annurev-astro-081309-130834},
  48, 431

\bibitem[\protect\citeauthoryear{Proszkow \& Adams}{Proszkow \&
  Adams}{2009}]{Proszkow2009}
Proszkow E.-M.,  Adams F.~C.,  2009, \mn@doi [The Astrophysical Journal
  Supplement Series] {10.1088/0067-0049/185/2/486}, 185, 486

\bibitem[\protect\citeauthoryear{Raboud \& Mermilliod}{Raboud \&
  Mermilliod}{1998}]{Raboud1998}
Raboud D.,  Mermilliod J.~C.,  1998, Astronomy and Astrophysics, 333, 897

\bibitem[\protect\citeauthoryear{Ranalli, Hobbs  \& Lindegren}{Ranalli
  et~al.}{2018}]{Ranalli2018}
Ranalli P.,  Hobbs D.,   Lindegren L.,  2018, \mn@doi [Astronomy {\&}
  Astrophysics] {10.1051/0004-6361/201730921}, 614, A30

\bibitem[\protect\citeauthoryear{Rice, Rasio  \& Steffen}{Rice
  et~al.}{2018}]{Rice2018}
Rice D.~R.,  Rasio F.~A.,   Steffen J.~H.,  2018, \mn@doi [Monthly Notices of
  the Royal Astronomical Society] {10.1093/mnras/sty2418}, 481, 2005

\bibitem[\protect\citeauthoryear{Scally \& Clarke}{Scally \&
  Clarke}{2001}]{Scally2001}
Scally A.,  Clarke C.,  2001, \mn@doi [Monthly Notices of the Royal
  Astronomical Society] {10.1046/j.1365-8711.2001.04274.x}, 325, 449

\bibitem[\protect\citeauthoryear{Shara, Hurley  \& Mardling}{Shara
  et~al.}{2016}]{Shara2016}
Shara M.~M.,  Hurley J.~R.,   Mardling R.~A.,  2016, \mn@doi [The Astrophysical
  Journal] {10.3847/0004-637X/816/2/59}, 816, 59

\bibitem[\protect\citeauthoryear{Spurzem, Giersz, Heggie  \& Lin}{Spurzem
  et~al.}{2009}]{Spurzem2009}
Spurzem R.,  Giersz M.,  Heggie D.~C.,   Lin D. N.~C.,  2009, \mn@doi
  [Astrophysical Journal] {10.1088/0004-637X/697/1/458}, 697, 458

\bibitem[\protect\citeauthoryear{Triaud}{Triaud}{2018}]{Triaud2018}
Triaud A. H. M.~J.,  2018, in Deeg H.~J.,  Belmonte J.~A.,  eds, , Handbook of
  Exoplanets.
Springer International Publishing, Cham, pp 1375--1401, \mn@doi{10.1007/978-3-319-30648-3_2-1}

\bibitem[\protect\citeauthoryear{Vincke \& Pfalzner}{Vincke \&
  Pfalzner}{2016}]{Vincke2016}
Vincke K.,  Pfalzner S.,  2016, \mn@doi [The Astrophysical Journal]
  {10.3847/0004-637X/828/1/48}, 828, 48

\bibitem[\protect\citeauthoryear{Winter, Clarke, Rosotti, Ih, Facchini  \&
  Haworth}{Winter et~al.}{2018}]{Winter2018}
Winter A.~J.,  Clarke C.~J.,  Rosotti G.,  Ih J.,  Facchini S.,   Haworth
  T.~J.,  2018, \mn@doi [Monthly Notices of the Royal Astronomical Society]
  {10.1093/mnras/sty984}, 478, 2700

\bibitem[\protect\citeauthoryear{Zhu \& Wu}{Zhu \& Wu}{2018}]{Zhu2018a}
Zhu W.,  Wu Y.,  2018, \mn@doi [The Astronomical Journal]
  {10.3847/1538-3881/aad22a}, 156, 92

\bibitem[\protect\citeauthoryear{Zhu, Petrovich, Wu, Dong  \& Xie}{Zhu
  et~al.}{2018}]{Zhu2018}
Zhu W.,  Petrovich C.,  Wu Y.,  Dong S.,   Xie J.,  2018, \mn@doi [The
  Astrophysical Journal] {10.3847/1538-4357/aac6d5}, 860, 101

\bibitem[\protect\citeauthoryear{van Elteren, {Portegies Zwart}, Pelupessy, Cai
   \& McMillan}{van Elteren et~al.}{2019}]{VanElteren2019}
van Elteren A.,  {Portegies Zwart} S.,  Pelupessy I.,  Cai M.~X.,   McMillan S.
  L.~W.,  2019, \mn@doi [Astronomy {\&} Astrophysics]
  {10.1051/0004-6361/201834641}, 624, A120

\makeatother
\end{thebibliography}

\bsp
\label{lastpage}
\end{document}